\documentclass[reprint,aps,amsmath,amssymb,superscriptaddress]{revtex4-1}

\usepackage{graphicx} 
\usepackage[utf8]{inputenc}
\usepackage{bm}
\usepackage{hyperref}
\usepackage{xcolor}
\usepackage[normalem]{ulem}
\usepackage{amsmath}
\usepackage{float}
\usepackage{algorithm}
\usepackage{algpseudocode}
\usepackage{amsmath}
\usepackage{comment}
\setcitestyle{authoryear,round}

\begin{document}

\title{Employment-access constraints in the 15-minute city}

\author{Marc Barthelemy}
\email{marc.barthelemy@ipht.fr}
\affiliation{Universit\'e Paris-Saclay, CNRS, CEA, Institut de Physique Th\'eorique,
91191 Gif-sur-Yvette, France}
\affiliation{Centre d'Analyse et de Math\'ematique Sociales (CNRS/EHESS), Paris, France}
\affiliation{Complexity Science Hub, Vienna, Austria}

\date{\today}

\begin{abstract}

The ``15-minute city'' has emerged as an influential urban-planning paradigm,
promoting proximity to work and essential services. Its feasibility for
journey-to-work commuting, however, has rarely been examined quantitatively. 
Here, we show that employment proximity is constrained by heterogeneity in the number of employees per firm. Combining urban geometry with empirically observed
firm-size distributions, we derive a lower bound on commuting time that holds
even under optimal workplace placement. This bound reveals a sharp transition: when employment is sufficiently unevenly distributed across firms, no spatial
rearrangement of workplaces can provide uniformly short commutes. Applied to
Paris and its near suburbs, with area \(A=762~\mathrm{km}^2\) and empirical
firm-size exponent \(\hat{\gamma}\simeq 1.38\), the model predicts that only
about \(37\%\) of workers can reach their workplace within \(15\) minutes by
walking. Cycling increases this share to about \(67\%\), but still leaves
roughly one third of the workforce beyond the 15-minute threshold. Achieving
universal short commutes would therefore require economic restructuring,
differentiated mobility strategies, remote work, or a longer target time. Our
results suggest replacing the universal 15-minute target, for commuting, by a
city-specific benchmark determined by firm-size heterogeneity and urban scale.

\end{abstract}

\maketitle

\section*{Main}

Building on earlier traditions of neighborhood-based urbanism and accessibility planning,
the ``15-minute city''~\citep{moreno2021} (or more generally the $x$-minute city) promotes
the idea that essential urban functions---work, commerce, healthcare, and
education---should be reachable within a short travel time from one's residence,
ideally by walking or cycling. Initially framed as a response to climate change, social
inequalities, and car-dependent urban growth, it has rapidly gained visibility as a
normative model for sustainable urban development~\citep{moreno2025}, promising reduced car
dependence and emissions alongside improvements in quality of life, public health, and
social inclusion~\citep{saelens2003}.

Although employment is explicitly included in the original formulation~\citep{moreno2021}, it has proven difficult to operationalize empirically, and commuting-to-work constraints have received limited quantitative attention. Concerns about job–housing proximity are longstanding: classic work on accessibility and job–housing balance already showed that residences in job-rich areas are associated with shorter commutes~\citep{levinson1998}. The 15-minute city can thus be viewed as a strengthening of this principle, aiming not merely at average reductions in
commuting time but at universal proximity for all workers. Recent empirical work,
however, shows that proximity to services does not necessarily translate into localized
travel, particularly for work trips~\citep{arcaute2025a}. 

Methodological reviews confirm this gap between conceptual ambition and empirical practice, showing that the 15-minute city lacks a unified operational definition, relies on heterogeneous indicators, and predominantly measures access to local services, while commuting responds more weakly to proximity-based interventions~\citep{buettner2024,omwamba2025,papadopoulos2023measuring,pozoukidou2021,khavarian2023,abbiasov202415,rhoads2023inclusive}. Large-scale empirical assessments further quantify substantial heterogeneity of accessibility within and across cities worldwide, highlighting the role of population density and resource distribution in shaping proximity outcomes~\citep{buettner2024,omwamba2025,bruno2024universal}. Despite these advances, the feasibility of the $x$-minute city remains poorly understood at a more fundamental level. In particular, recent reviews~\citep{buettner2024,omwamba2025} make clear that most studies stop short of addressing a structural question: whether universal proximity to work is achievable \emph{in principle}, independently of specific planning choices, service redistribution strategies, or mobility technologies. In particular, \citet{bruno2024universal} propose a global framework to quantify accessibility to essential services and to simulate their optimal redistribution across cities worldwide. They measure proximity times to nine categories of point-of-interests (outdoor activities, learning, supplies, eating, moving, cultural activities, physical exercise, services, and healthcare), and develop an algorithm to relocate these amenities so as to equalise services per capita and increase the fraction of the population living within a 15-minute condition. Their results highlight strong heterogeneity across cities and show that substantial service redistribution would be required in many urban contexts. 

However, the most significant structural obstacle to achieving a universal 15-minute city lies in commuting to work. Unlike local amenities such as shops or schools, employment is intrinsically heterogeneous: a small number of large firms employ a substantial fraction of the workforce, while many small establishments coexist alongside them. Crucially, employment is not a freely replicable service that can be redistributed without altering the underlying economic structure itself. This difference is therefore not merely quantitative but structural. More precisely, firm sizes follow broad, approximately Zipfian distributions across countries, sectors, and time periods~\citep{ramsden2000company,axtell2001,gaffeo2003,gabaix2016,santos2024zipf}. Although firm-level employment data are typically available only in aggregated form at
the urban scale, establishment counts reveal substantial within-city heterogeneity, with
large employers accounting for a disproportionate share of
employment (see section~I of the SI). In general, we can expect that large metropolitan areas necessarily host the upper tail of the firm-size
distribution~\citep{sridhar2010firm}, making economic heterogeneity a structural
constraint at the urban scale.

This observation raises a fundamental question left open by existing
studies~\citep{buettner2024,omwamba2025}: under what conditions is it possible, even in
principle, to locate firms in space such that every worker can reach their workplace
within a prescribed travel time $\tau_0$? We address this question using a minimal
spatial framework combining three ingredients: (i)~a population distributed over a city
of area~$A$, (ii)~firms whose sizes follow a Zipf distribution with exponent~$\gamma$,
and (iii)~an optimization problem in which firm locations are chosen to minimize the maximum
commuting distance. This framework yields strict theoretical bounds on proximity, which
we confront with numerical optimization and a case study of Paris and its near suburbs.

\section*{Model}

To isolate the role of the journey-to-work commute, we introduce a minimal model that focuses on this single ingredient and reveals the existence of a structural lower bound on commuting time. We thus consider an idealized city of characteristic radius $R$ (typically of order
$\sqrt{A/\pi}$, where $A$ is the urban area), and total employment $E$. Residents are assumed to be uniformly distributed in space unless stated otherwise. Employment is provided by $N$ firms whose sizes are heterogeneous and ranked by decreasing employment $m_1>m_2>\dots>m_N$. Firm sizes are assumed to follow a Zipf distribution, such that the firm of rank $r$ employs $m_r \propto r^{-\gamma}$ workers, where the exponent $\gamma$ controls the
degree of economic inequality: $\gamma=0$ corresponds to homogeneous firms, while larger values of $\gamma$ indicate increasingly concentrated employment in a small number of large firms. Workers are then assigned to firms subject to their capacity constraints. In the most optimistic version of the model, each worker is assigned to the nearest available firm, so that commuting distances are minimized as much as possible
given the spatial distribution and capacities of firms. The model therefore provides a best-case benchmark for commuting: it ignores all constraints other
than firm capacity. Real labour markets involve skill matching, wages, contracts,
individual preferences, sectoral specialization, and firm clustering, all of which restrict worker--firm matching further and would generally increase commuting distances.

This framework is intentionally minimal and aims to identify feasibility constraints that are independent of detailed urban design, land-use regulation, or transport technologies. It applies most directly at spatial scales where employment locations compete for access to the same pool of residents, such as neighborhoods, districts, or central urban areas.
At larger metropolitan scales, many cities exhibit a polycentric organization with multiple employment centers rather than a single dominant core~\citep{anas1998,berroir2011mobility,bertaud2003spatial,lenechet2012urban,loubar2014}. In such cases, proximity constraints should be evaluated at the level of individual subcenters rather than across the entire metropolitan area. Applying a 15-minute constraint at the metropolitan scale is therefore not meaningful in general, as commuting distances necessarily exceed this threshold even under optimistic mobility assumptions. Thus, proximity-based planning implicitly assumes a decomposition of cities into smaller functional units, and the present framework provides a feasibility test for such units.

The normalization condition $\sum_{r=1}^N m_r = E$, together with $m_r = m_1 r^{-\gamma}$, fixes the size of the largest firm as
\begin{equation}
m_1 = \frac{E}{H_N(\gamma)},
\label{eq:m1}
\end{equation}
where $H_N(\gamma)=\sum_{r=1}^N r^{-\gamma}$ is the $n$-th generalized harmonic number. For large $N$,
$H_N(\gamma)\sim N^{1-\gamma}/(1-\gamma)$ for $\gamma<1$, $H_N(1)\sim\log N$ in the marginal
case ($\gamma=1$), and $H_N(\gamma)\to\zeta(\gamma)$ for $\gamma>1$, where $\zeta(\gamma)$ denotes the
Riemann zeta function. This result Eq.~\ref{eq:m1} shows that the firm-size distribution alone determines how many workers must be served by the largest employer, independently of spatial considerations. 

Each worker is assigned to a single firm, and commuting is approximated by straight-line (Euclidean) distances, a standard first-order approximation in  urban spatial models \citep{batty2013}, which captures the underlying spatial constraint independently of network details. For a given spatial arrangement of firms, we define $\ell_{\max}$ as the maximum commuting distance across the population. The central feasibility question is whether firms can be located in space such that
$\ell_{\max}\le L_0$, where $L_0$ denotes a prescribed maximum commuting distance (or, equivalently, a maximum commuting time). Geometrically, the worker--firm assignment problem is equivalent to a weighted Voronoi partition~\citep{okabe2009spatial}, in which firm sizes act as weights constraining the spatial extent of their catchment areas. As firm-size
heterogeneity increases, these weighted cells become increasingly distorted, forcing some regions to span large distances and driving the proximity feasibility transition.

\section*{Results}

\subsection*{Theoretical feasibility constraints}
\label{subsec:theory}

We derive a strict theoretical constraint on the feasibility of an $x$-minute city that holds independently of any numerical optimization. This constraint follows from elementary geometric considerations combined with firm-size heterogeneity and defines an absolute lower bound that no spatial rearrangement of workplaces can overcome.

The main idea is that the dominant constraint is imposed by the largest firm. Assuming that firm sizes follow a Zipf distribution with exponent $\gamma$, the largest firm employs a fraction of the total workforce given by Eq.~\ref{eq:m1}. Even if this firm is optimally located at the point of
maximum population density, it must recruit its workforce from a finite catchment area which has a certain size. For
a uniform worker density $\rho=E/A$, the radius $\ell_{\max}$ of this catchment area
satisfies $\rho \pi \ell_{\max}^2 = m_1$. Requiring that all workers commute less than a
prescribed distance $L_0$ therefore yields the lower bound
\begin{equation}
L_0 \;\ge\; L_0^{\ast} \equiv \sqrt{\frac{A/\pi}{H_N(\gamma)}} .
\label{eq:fund}
\end{equation}
In the numerical model on a disk of
radius $R$, this reduces to $L_0/R \ge 1/\sqrt{H_N(\gamma)}$. Expressed in terms of time rather than distance, the minimal commuting time achievable in
this city is then
\begin{equation}
\tau_0(\gamma)
\;=\;
\frac{1}{v}\,\frac{\sqrt{A/\pi}}{\sqrt{H_N(\gamma)}},
\label{eq:tau_min_general}
\end{equation}
where $v$ is the characteristic travel speed (walking or biking). Equations~(\ref{eq:fund},\ref{eq:tau_min_general}) constitute a fundamental limit on urban proximity: it cannot be improved by redistributing firms, modifying transport infrastructure, or altering land-use patterns. It provides the
reference against which numerical optimizations and empirical case studies are interpreted below (more details and discussion can be found in section II of the SI).

The factor $\sqrt{A/\pi}/v$ corresponds to the time required to traverse the city radius, while the effect of firm-size heterogeneity enters through the denominator $\sqrt{H_N(\gamma)}$. The resulting constraint depends qualitatively on the value of $\gamma$. For $\gamma>1$, the harmonic number converges to $H_N(\gamma)\to\zeta(\gamma)$ in
the large-$N$ limit, yielding a finite asymptotic lower bound
$L_0^{\ast}=\sqrt{A/\pi}/\sqrt{\zeta(\gamma)}$. In this regime, increasing the number of firms cannot reduce commuting distances below a fixed fraction of the city size, revealing a structural obstruction to universal short commutes. By contrast, in the homogeneous limit $\gamma=0$, one has $H_N(0)=N$ and the bound reduces to the purely geometric scaling $L_0^{\ast}=\sqrt{A/\pi}/\sqrt{N}$. This quantity can be made arbitrarily small by increasing the number of firms,
subject to the natural constraint that the number of firms cannot exceed the number of employees, \(N \leq E\).

The transition between these regimes reflects a qualitative change in the nature of proximity constraints, from geometry-dominated at low heterogeneity to extreme-value-dominated at large $\gamma$. Empirically, estimates based on binned establishment-size data for U.S. metropolitan areas yield values in the range $\gamma\in[0.4,1.0]$, with mean $\langle\gamma\rangle\simeq0.71\pm0.12$ (see the section I of the SI). For the Paris metropolitan area, available data are consistent with a larger exponent, $\gamma\simeq1.4$ (see section V of the SI), placing it deep in the high-heterogeneity regime where proximity constraints become intrinsically difficult to satisfy at the metropolitan scale.

\subsection*{Numerical phase diagram}
\label{subsec:numerical}

We next examine how the theoretical feasibility bounds manifest at finite system size using explicit spatial optimization. The aim of this numerical analysis is not to calibrate real cities, but to probe the robustness and sharpness of the feasibility constraint with finite numbers of workers and firms.

\begin{figure*}
  \centering
  \includegraphics[width=0.9\textwidth]{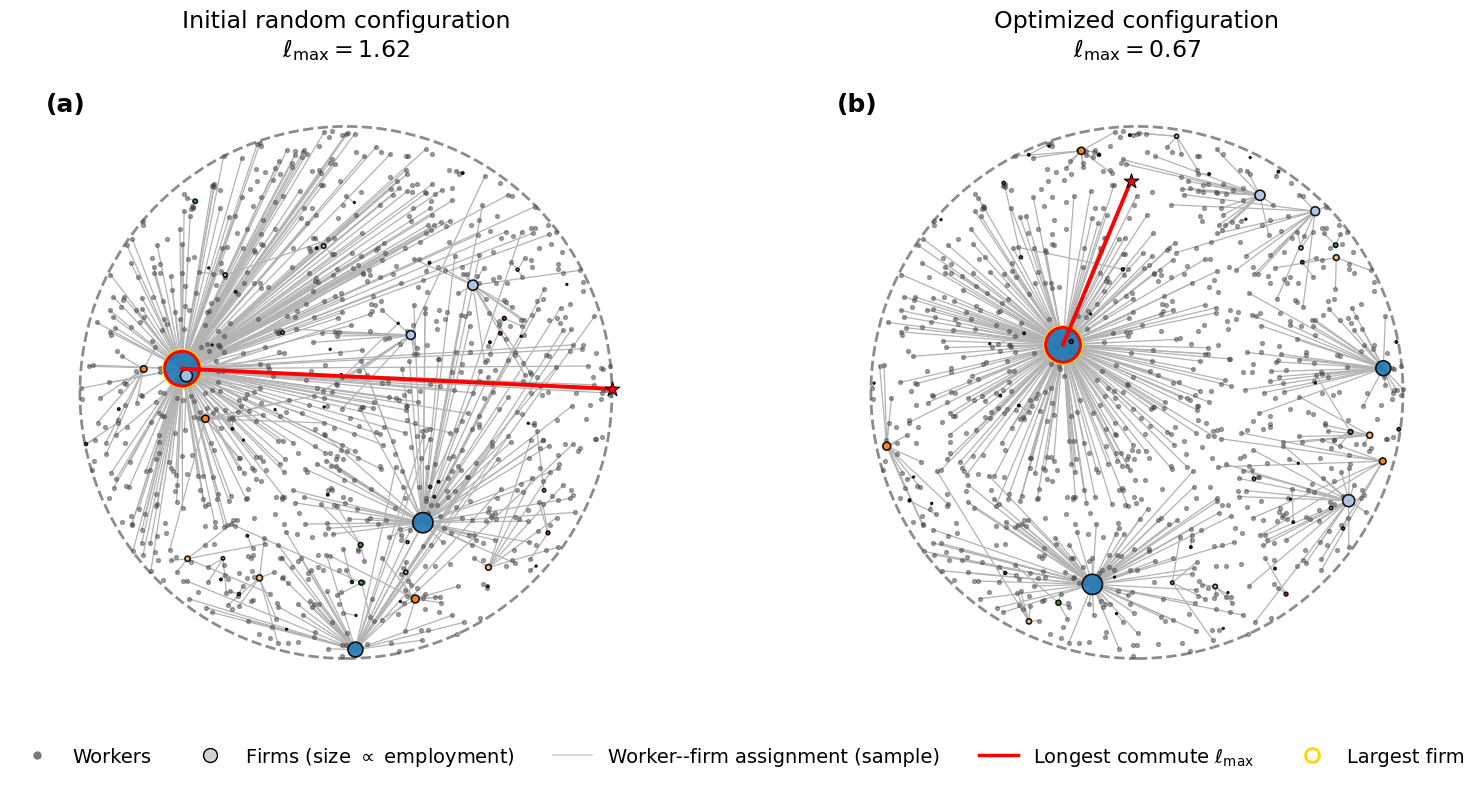}
\caption{
\textbf{Relocation and assignment optimization.}
Representative simulated annealing example with \(E=1000\) workers and
\(N=60\) firms, whose sizes follow \(m_n\propto n^{-\gamma}\) with
\(\gamma=1.5\) (\(m_1=424.6\), \(m_N=1\)). Workers are shown as small points
and firms as circles with area proportional to employment capacity. For visual
clarity, assignment links are shown only for a sample of workers, although the
full worker--firm assignment is used in the optimization. The highlighted link
indicates the longest commute \(\ell_{\max}\).
(a) Initial random firm configuration, with
\(\ell_{\max}\simeq 1.62\).
(b) Optimized configuration after simulated annealing
(\(T_{\mathrm{init}}=0.5\), \(T_{\min}=10^{-4}\), \(\alpha=0.95\),
\(n_{\mathrm{steps}}=1000\) per temperature), with
\(\ell_{\max}\simeq 0.67\), corresponding to a reduction of approximately
\(59\%\). The optimization relocates the largest firms toward the centre of the
population distribution, reducing the extreme distances that control
\(\ell_{\max}\).
}
  \label{fig:illus}
\end{figure*}

\begin{figure*}
  \centering
  \includegraphics[width=0.9\textwidth]{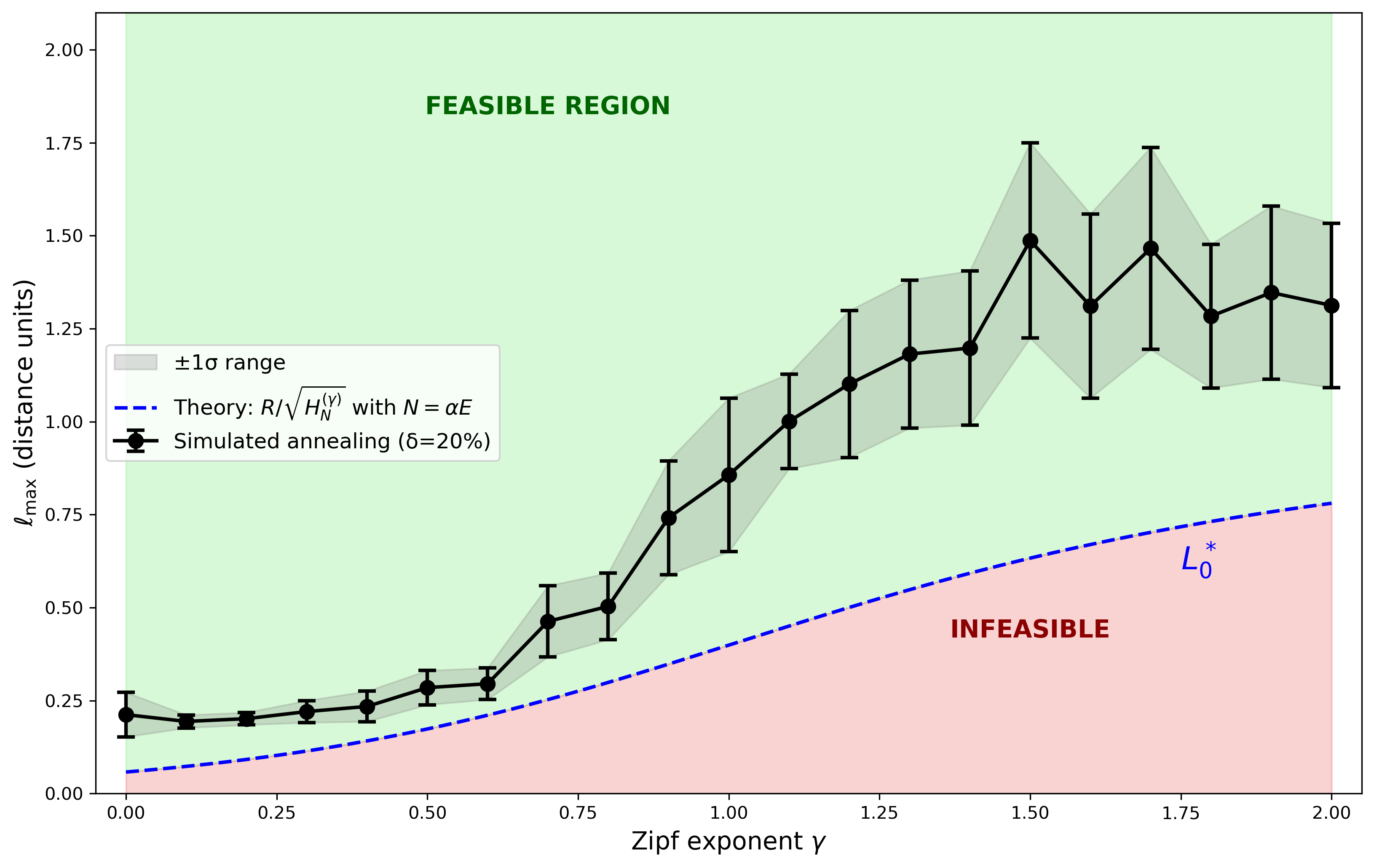}
  \caption{
  \textbf{Finite-size phase diagram of proximity feasibility.}
  Maximum commuting distance $\ell_{\max}$ obtained by simulated annealing as a function of the firm-size exponent $\gamma$. Black circles show averages over 10 independent spatial configurations, and error bars indicate one standard deviation. Simulations are performed for total employment $E=5000$, city radius $R=1$, number of establishments $N=\alpha E$ with $\alpha=0.06$, and capacity tolerance $\delta=20\%$. The dashed line indicates the theoretical lower bound $L_0^{\ast}$ given by Eq.~\ref{eq:fund} for a uniform population density.
  The region below this bound is rigorously infeasible: no spatial
  arrangement of establishments can achieve a smaller maximum commuting
  distance. The region above the bound is not excluded by the analytical
  constraint, and the simulated-annealing solutions provide feasible
  numerical upper estimates. The gap between the symbols and the lower bound
  may reflect finite-size and discreteness effects, geometric constraints, and
  possible suboptimality of the annealing procedure.
}
  \label{fig:phase_diagram}
\end{figure*}

For fixed values of $(\gamma,E,N)$ and a prescribed capacity tolerance $\delta$, firm locations are optimized using simulated annealing~\citep{kirkpatrick1983} to minimize the maximum commuting distance $\ell_{\max}$. The tolerance $\delta$ quantifies the allowed mismatch between
firm capacity and the number of workers assigned to it: a firm of size $m$ can accommodate a number of workers in the interval $[(1-\delta)m,(1+\delta)m]$. This parameter captures realistic flexibility in firm capacity, congestion effects, and imperfect matching between workers and
establishments. As derived in Section~III of the SI, capacity flexibility modifies the theoretical lower bound by replacing the factor $1/H_N(\gamma)$ with $(1-\delta)/H_N(\gamma)$, yielding a reduced minimum commuting time
$\tau_0(\gamma,\delta)
  = \tau_0(\gamma)\,\sqrt{1-\delta}$. The optimization is repeated over multiple independent realizations of the
population distribution to estimate both the mean optimal $\ell_{\max}$ and its
fluctuations (for more details on the numerical simulation, see the section IV in the SI). Throughout this analysis, we assume that the number of establishments scales proportionally with total employment, $N\simeq\alpha E$, following empirical observations for U.S. Metropolitan Statistical Areas, with $\alpha\simeq0.06$. This assumption fixes the average firm size, $\langle m\rangle=1/\alpha$, and closes the model (see the SI). Because the capacitated assignment is solved by a sequential greedy heuristic,
the numerical value of \(\ell_{\max}\) can depend on the worker processing
order. As shown in the SI (section IV-C), repeating the full simulated annealing with fixed
random orderings gives the same qualitative reduction of \(\ell_{\max}\), but
larger final values, confirming that the numerical configurations are feasible
heuristic upper estimates rather than proven global optima.

We illustrate the optimization procedure in Fig.~\ref{fig:illus} for a
representative example with \(E=1000\) workers and \(N=60\) firms, whose sizes are distributed according to a Zipf law with exponent \(\gamma=1.5\). Starting from random firm locations, workers are assigned to firms subject to the capacity constraints. The simulated annealing algorithm then relocates firms and recomputes the assignments in order to minimize the maximal commuting distance \(\ell_{\max}\). In the initial configuration, the maximal commute is
\(\ell_{\max}\simeq 1.62\), close to the diameter \(2\) of the disk. After
optimization, the maximal commuting distance decreases to
\(\ell_{\max}\simeq 0.67\), corresponding to a reduction of about \(59\%\), or
roughly one third of the system diameter. The optimized configuration exhibits
a clear spatial organization: the largest firms are relocated toward the
centre of the population distribution, while smaller firms are distributed
closer to the periphery. This arrangement reduces extreme commuting distances
and prevents the occurrence of very long trips. The example illustrates the
basic mechanism behind the analytical bound: because large firms require large
catchment areas, their position has a disproportionate effect on
\(\ell_{\max}\).

The previous example illustrates how spatial optimization can reduce extreme commuting distances in a single configuration. We now examine how this limit depends on the degree of economic heterogeneity. Figure~\ref{fig:phase_diagram} shows the maximal commuting distance $\ell_{\max}$ obtained by simulated annealing as a function of the firm-size exponent $\gamma$, together with the theoretical lower bound $L_0^{\ast}$. Two main results emerge. First, the numerical solutions reproduce the qualitative trend predicted by the theory: increasing firm-size heterogeneity leads to a monotonic increase in the best commuting distance found by the algorithm. The change from feasible to infeasible proximity regimes is therefore consistent with the theoretical prediction that heterogeneity acts as the dominant control parameter. Second, the numerical configurations lie systematically above the theoretical bound. This gap should be interpreted with caution. It may reflect finite-size effects, the discrete nature of the spatial optimization problem, geometric constraints, and possible suboptimality of the simulated-annealing procedure. While the analytical result provides an asymptotic lower bound, numerical realizations involve finite numbers of firms and residents and are obtained by a heuristic optimization algorithm. The theoretical curve therefore acts as a strict lower bound, whereas the annealed configurations provide feasible numerical upper estimates at moderate system size.

The discrepancy between numerical and theoretical bounds becomes large around and above $\gamma\simeq1$, a range consistent with empirically observed Zipf-like firm-size distributions. This value marks a qualitative crossover between weakly and strongly heterogeneous regimes. For $\gamma>1$, firm sizes span
several orders of magnitude without a characteristic scale, and the optimization becomes dominated by a small number of extreme firms. In this regime, $\ell_{\max}$ is controlled by rare, system-spanning constraints rather than by
typical firms, leading to an abrupt breakdown of proximity.

Overall, these numerical results show that the infeasibility transition predicted by the theory is already visible at moderate system sizes. 
In finite systems, proximity constraints are even stronger: the optimized maximum commuting distance is systematically larger than the theoretical lower bound.

\subsection*{Paris as a stress test for the 15-minute city}
\label{subsec:paris_case}

Paris provides a particularly challenging case for the feasibility of a 15-minute city. Approximately $1.8$ million individuals work inside the city of Paris, but commuting flows are highly asymmetric: nearly $60\%$ ($1{,}080{,}000$ individuals) of workers employed in Paris reside outside the city (data for 2017;~\citep{insee_paris_navetteurs}). Among these $1.08$ million workers, about $55\%$ come from the near suburbs (the \emph{petite couronne}), with the remainder residing in more distant suburbs and other regions. We therefore focus on the $1.314$ million individuals who work in Paris and live either in Paris or in its near suburbs, excluding the outer suburbs (the \emph{grande couronne}), where commuting distances typically exceed $20$~km and a 15-minute constraint is unrealistic.

As an empirical benchmark, observed home--work commuting times in this area
are already substantially larger than the 15-minute threshold. Based on the
2008 French National Transport Survey \citep{Caenen2011FranciliensTravail}, a mean commuting time of $31$ minutes for workers living in Paris and
$33$ minutes for workers living in the \emph{petite couronne} were reported; for
$\mathrm{\hat I le}$-de-France as a whole, the corresponding value is
$34$ minutes. These averages include all transport modes, in particular
public transport and car trips. The same source reports that public transport
accounts for $64.0\%$ of home--work trips in Paris and $44.8\%$ in the
\emph{petite couronne}, while car trips account for $13.3\%$ and $38.0\%$,
respectively. Thus, the observed mean commuting time in the area considered
here is roughly twice the 15-minute threshold.

Paris hosts approximately $406\,000$ establishments~\citep{greffe_paris},
corresponding to an average firm size of
$\langle m\rangle\simeq4.4$~employees.
Rather than relying on a single estimate of $N$ for the combined
area of Paris and the \emph{petite couronne} (which for example would be $\sim 3\times 10^5$ for $4.4$ employees per firm), we vary $N$ over almost two
orders of magnitude, from $10^4$ to $5\times10^5$, encompassing any plausible
value. To characterize firm-size heterogeneity, we estimate the Zipf exponent $\gamma$ using establishment-size data available at the scale of the Paris urban area. Using INSEE data
for 2023~\citep{insee_2023}, we obtain a relatively large value $\gamma\simeq1.38$ (see section V of the SI),
indicating strong employment concentration. Given uncertainties in the estimation, we treat $\gamma$ as a control parameter and explore a broad range of values.

The central question is whether these $N$ (with $N\in [10^4,5\times10^5]$) establishments can be
spatially reorganized over Paris and the \emph{petite couronne} such that all
$1.314$ million workers commute less than 15 minutes. To address this question, we use the
theoretical lower bound $\tau_0(\gamma)$ derived above (Eq.~\eqref{eq:tau_min_general}). At this spatial scale, we approximate the population density as uniform, which
allows us to apply the analytical framework directly. This approximation is
supported by the empirical results of \citet{lemoy2020evidence}, who find that
land-use density is approximately constant up to distances
\(r\simeq 20\,\mathrm{km}\) from the city centre, a range that includes Paris
and the \emph{petite couronne} (the exponential-density calculation developed in the SI shows how the bound would change under stronger central
concentration - which is less appropriate for the Paris--\emph{petite couronne} scale considered here).

\begin{figure*}[t]
  \centering

  \newcommand{\mapheight}{0.46\textheight}
  \newcommand{\rightfigheight}{0.18\textheight}

  \begin{minipage}[t]{0.46\textwidth}
    \vspace{0pt}
    \centering
    \includegraphics[
      width=\linewidth,
      height=\mapheight,
      keepaspectratio
    ]{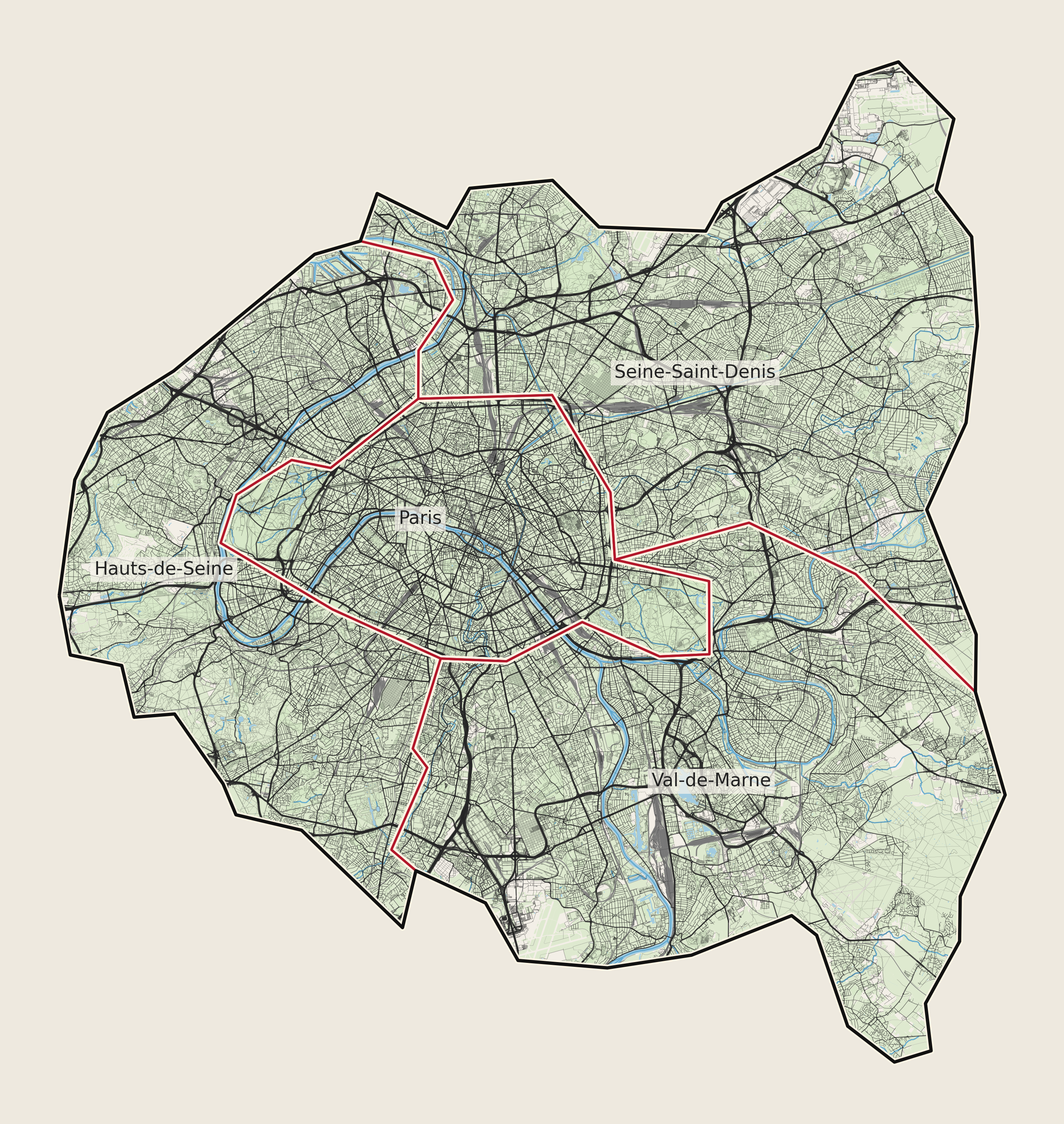}
  \end{minipage}
  \hfill
  \begin{minipage}[t]{0.51\textwidth}
    \vspace{0pt}
    \centering

    \includegraphics[
      width=\linewidth,
      height=\rightfigheight,
      keepaspectratio
    ]{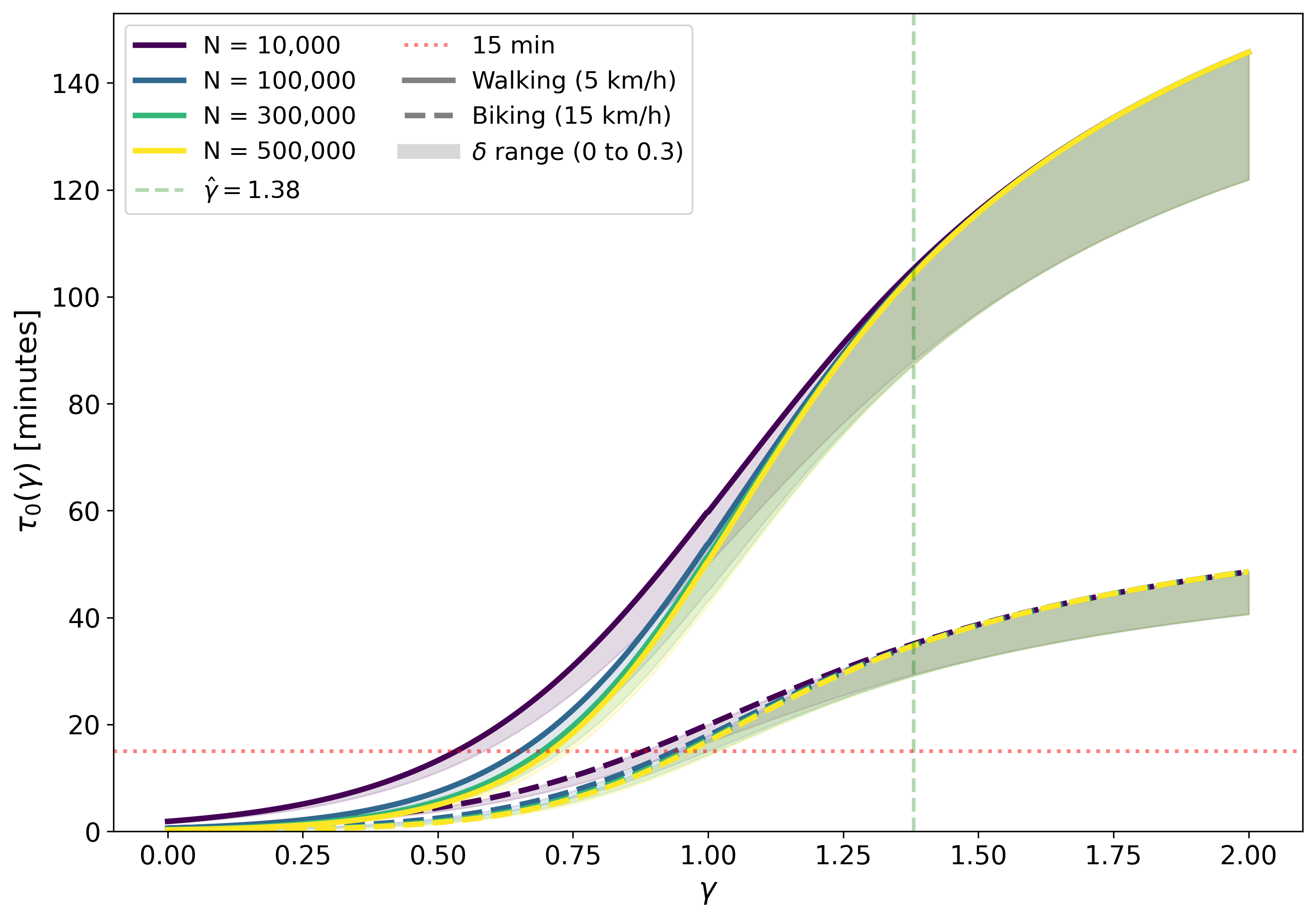}

  \vspace{0.8em}
  
  \includegraphics[
      width=\linewidth,
      height=\rightfigheight,
      keepaspectratio
    ]{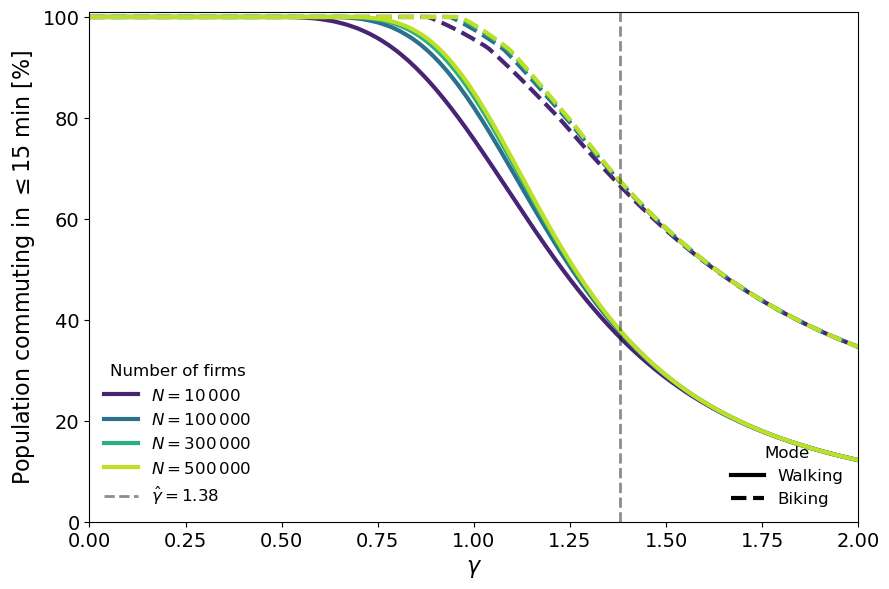}

  \end{minipage}

  \caption{
  Accessibility constraints in Paris and the \emph{petite couronne}.
  (a) Map of Paris and the \emph{petite couronne}---Paris,
  Hauts-de-Seine, Seine-Saint-Denis, and Val-de-Marne---showing the road
  network, water bodies, green areas, railways, and department boundaries.
  Map data are from OpenStreetMap contributors, distributed by Geofabrik under
  the Open Database License (ODbL). Department boundaries are from France
  GeoJSON, based on official French administrative data. (b) Minimal achievable commuting time \(\tau_0(\gamma)\) as a function of
\(\gamma\), computed for the same area. Solid curves correspond to walking and
dashed curves to cycling; colour encodes the number of establishments \(N\).
When \(N\) is varied, the average number of employees per establishment is kept
constant, \(E/N=1/\alpha\simeq 4.4\), so that the total workforce \(E\) varies
proportionally with \(N\). The shaded band below each curve shows the reduction
in \(\tau_0\) allowed by a capacity flexibility \(\delta\in[0,0.3]\). The
horizontal dotted line marks the 15-minute threshold, and the vertical dashed
line indicates \(\hat{\gamma}\simeq1.38\). (c) Share of the workforce able to reach work within \(15\) minutes,
\(P_{15}(\gamma)\), for the Paris urban area
(\(A=762~\mathrm{km}^2\)), as a function of the firm-size exponent
\(\gamma\). Colours indicate the number of establishments \(N\), with
\(N=10^4\), \(10^5\), \(3\times 10^5\), and \(5\times 10^5\). Solid curves
correspond to walking (\(5~\mathrm{km/h}\)) and dashed curves to cycling
(\(15~\mathrm{km/h}\)). The vertical dashed line marks the empirical exponent
\(\hat{\gamma}\simeq 1.38\), for which the accessible fraction varies only
weakly with \(N\) ($\approx 37-38\%$) for walking and about $67\%$ for cycling.}
  \label{fig:paris_accessibility_combined}
\end{figure*}

Figure~\ref{fig:paris_accessibility_combined}(a) shows the spatial domain under consideration: Paris and the \emph{petite couronne}. Panel~(b) shows the theoretical minimum
commuting time \(\tau_0(\gamma)\) as a function of firm-size heterogeneity,
for both walking and cycling, and for a wide range of establishment counts $N$. Two regimes emerge. For \(\gamma>1\), the curves obtained for
different values of $N$ collapse onto a single curve. In this
regime, \(H_N(\gamma)\) is well approximated by \(\zeta(\gamma)\), so that the
lower bound depends only on the urban area, the travel speed, and the
firm-size exponent, but not on the total number of establishments. For
\(\gamma<1\), by contrast, \(H_N(\gamma)\) increases algebraically with $N$ (and logarithmically at \(\gamma=1\)), making \(\tau_0\)
sensitive to the precise number of establishments. In this weakly
heterogeneous regime, increasing $N$ distributes workers among
more, smaller catchments and therefore reduces the maximum commuting distance.

The empirical firm-size distribution for Paris, however, lies in the
\(\gamma>1\) regime. The theoretical bound is therefore essentially
independent of the number of establishments. The shaded bands quantify the
additional gain allowed by capacity flexibility \(\delta\). This correction
acts through the multiplicative factor \(\sqrt{1-\delta}\) (see SI), which is
independent of \(\gamma\). Even for \(\delta=0.3\), corresponding to a
reduction of about \(16\%\) in \(\tau_0\), the correction remains too small to
bring commuting times below the 15-minute threshold for empirically relevant
values of \(\gamma\). In particular, at the empirical estimate
\(\hat{\gamma}\simeq1.38\), the threshold cannot be reached either by
increasing the number of establishments or by allowing substantial capacity
flexibility. Cycling significantly lowers commuting times, but it does not
fully compensate for the constraint imposed by firm-size heterogeneity. At the
scale of Paris and its near suburbs, these results indicate that achieving a
15-minute city for commuting would require not only faster mobility but also a
substantial reorganisation of economic activity. This conclusion is consistent with observed commuting times in Paris and the \emph{petite couronne}, which are already around \(31\)--\(33\) minutes on average when all transport modes are included \citep{Caenen2011FranciliensTravail}.

A single commuting-time threshold hides how long commutes are distributed among
workers. We therefore define $P_{15}(\gamma)$, the fraction of workers whose
commute is shorter than 15 minutes. A firm of rank $n$ employs a share of the total workforce \(E\) given by $p_n = m_n/E= n^{-\gamma}/H_N(\gamma)$. Assuming a uniform residential density \(E/A\), this firm must draw its workers from a catchment disk of radius \(\ell_n\)
defined by $\pi \ell_n^2 E/A= m_n$. Equivalently,
\[
  \ell_n = R\sqrt{p_n},
\]
where \(R=\sqrt{A/\pi}\) is the radius of a disk with the same area as the urban region.
The largest firm sets the lower-bound commuting distance \(\ell_{\max}=\ell_1\), and hence
the lower-bound commuting time $\tau_0=\ell_1/v$. For a time budget \(\tau=15\) minutes, the reachable distance is $r = v\tau$. Treating each catchment as a uniform disk, the fraction of firm \(n\)'s own
workers who can reach their workplace within the 15-minute budget is
\[
  \min\!\left(1,\frac{r^2}{\ell_n^2}\right)
\]
Since firm \(n\) contains a share \(p_n\) of the total workforce, its
contribution to the total workforce share served within 15 minutes is
\[
  p_n\min\!\left(1,\frac{r^2}{\ell_n^2}\right)
  =
  \min\!\left[
    p_n,
    \left(\frac{r}{R}\right)^2
  \right].
\]
Summing over all establishments gives
\begin{equation}
  P_{15}(\gamma)
  =
  \sum_{n=1}^{N}
  \min\!\left[
    p_n,
    \left(\frac{r}{R}\right)^2
  \right].
  \label{eq:P15_gamma}
\end{equation}
the share of workers able to commute in less than 15 minutes to their working place. Large firms have catchment radii larger than the distance reachable within 15 minutes. Each such firm contributes only the fixed amount \((r/R)^2\) to the total served workforce share, whereas smaller firms, whose catchments fit inside the 15-minute radius, contribute their full employment share \(p_n\).

The consequence is shown in Fig.~\ref{fig:paris_accessibility_combined}(c).
When firm sizes are nearly homogeneous, corresponding to small \(\gamma\), the
entire workforce can be served within the 15-minute budget for the range of
establishment numbers considered. As firm-size heterogeneity increases, the
served workforce share decreases, driven by the large catchments required by
the dominant employers. At the empirical exponent
\(\hat{\gamma}\simeq 1.38\), the accessible fraction depends only weakly on
\(N\): for \(N=10^4\) to \(5\times 10^5\), it varies from \(36.5\%\) to
\(37.9\%\) for walking and from \(66.6\%\) to \(67.6\%\) for cycling.
Thus, even when cycling is possible, roughly one third of the workforce remains
beyond the 15-minute threshold. The same mechanism that sets the lower bound \(\tau_0\) also determines how
long commutes are distributed among workers.

\section*{Structural limits to urban proximity}

Our results identify a structural limitation of proximity-based urban ideals that
arises from the organization of employment rather than from transport
infrastructure alone. The key mechanism is firm-size heterogeneity. When
employment is distributed among firms of very different sizes, the largest
establishments must draw workers from spatially extended catchment areas in
order to fill their capacity. These large employment units therefore impose long
commuting distances even under optimal spatial placement. By contrast, when
employment is distributed among many small and relatively homogeneous firms,
the city can be decomposed into local catchment areas, allowing commuting
distances to remain short.

This result should not be interpreted as a failure of the 15-minute city as a
general planning principle. Rather, it identifies employment access as a
hard barrier to overcome. Many everyday services and amenities can, at least in
principle, be decentralized, whereas jobs are often distributed in a highly
heterogeneous and spatially concentrated way. Short commutes for
all workers cannot then be guaranteed solely by rearranging workplaces, even under
optimistic assumptions about travel speeds and worker assignment. The purpose
of the model is therefore not to provide a complete description of urban
complexity, but to isolate a theoretical constraint that arises from the
combination of urban geometry and the firm-size distribution.

The framework also provides a practical way to estimate the minimum achievable
proximity time in a given city. If empirical data are available for the
firm-size distribution, through $\gamma$, $N$, and $E$, and for the spatial
extent $A$ of the urban area, the theory gives a lower bound on commuting time
that cannot be overcome by spatial reorganization of workplaces alone. The
relevant question is therefore not only whether a city can be transformed into
an ``$x$-minute city'' for a prescribed value of $x$, but what the smallest
feasible value of $x$ is, given its economic structure and geometry. This
suggests a differentiated interpretation of proximity planning: local access to
daily amenities may be improved through decentralization, whereas employment
access in heterogeneous urban economies may require complementary mobility
solutions and efficient transport networks.

The model should also be understood as an optimistic benchmark. It treats
employment locations as units with a given capacity and assigns workers to the
nearest available unit. In real cities, firms are not placed independently in
space. Agglomeration economies favour clustering through shared labour markets,
supplier networks, knowledge spillovers, and sectoral specialization
\citep{marshall1961principles,duranton2004micro,rosenthal2004evidence}. The capacity parameter can therefore also be interpreted as the employment capacity
of a workplace cluster, business district, or specialized employment zone,
rather than as the capacity of a single isolated firm. Moreover, workers do not simply commute to the nearest available job, but to
jobs compatible with their skills, sector, wages, and career constraints. Explicitly modelling
agglomeration and worker--job matching would likely increase commuting
distances relative to the present idealized assignment rule. This lower bound is therefore a favourable benchmark for proximity. If short
commutes are difficult even in this idealized case, firm clustering and matching
constraints should make them even more difficult.

A related limitation concerns the treatment of population density. The main
results use a two-dimensional urban domain with uniform population density,
which provides a transparent baseline. Stronger central concentration, or
explicitly three-dimensional urban forms, could reduce the horizontal catchment
radius needed to supply workers to large firms and could therefore improve the
feasibility of a 15-minute commuting constraint. However, such extensions would
require additional assumptions about vertical travel time, building capacity,
land use, and worker--firm matching. An interesting inverse problem would be to
determine, for a given firm-size exponent $\gamma$, the density profile
required to satisfy a prescribed commuting-time threshold.

From a policy perspective, the constraint identified here could be mitigated by
measures that reduce excessive employment concentration, such as zoning limits
on highly concentrated office districts, incentives for secondary employment
centres, mixed-use planning that brings jobs closer to residential areas, and
the development of remote or hybrid working arrangements. In particular, remote work removes the commuting constraint for jobs that can be performed at a distance. For these workers, employment access is no longer limited by spatial proximity to the workplace, even when the firm-size distribution remains broad. Such policies would bring jobs closer to residents, but they must also be balanced against the benefits of firm clustering, including agglomeration economies and labour-market matching. The model therefore highlights a structural trade-off: local accessibility can be improved by decentralising employment or reducing the need for daily commuting, but this may conflict with economic forces that favour large firms and employment clusters.\\

{\bf Acknowledgments}: I thank Jieun Lee and Chris Webster for interesting discussions on the 15 minute city.\\

{\bf Data availability statement}: The datasets used in this study are freely available from public repositories: \citep{cbp_census,insee_paris_navetteurs,greffe_paris,insee_2023}.\\

{\bf Competing interests}: None.


\bibliography{ref_xmc}
\bibliographystyle{dcu}

\end{document}


\title{Supplementary Information\\ Employment-access constraints in the 15-minute city}


\author{Marc Barthelemy}
\email{marc.barthelemy@ipht.fr}
\affiliation{Université Paris-Saclay, CNRS, CEA, Institut de Physique Théorique,
91191 Gif-sur-Yvette, France}
\affiliation{Centre d'Analyse et de Mathématique Sociales (CNRS/EHESS),
54 Avenue de Raspail, 75006 Paris, France}
\affiliation{Complexity Science Hub,
Metternichgasse 8, 1030 Vienna, Austria}

\date{\today}

\maketitle

--------------------------------------------------
\section{Empirical analysis: Metropolitan Statistical Areas (US)}
\label{sec:empirical}

We present an empirical analysis of firm-size statistics for U.S.Metropolitan Statistical Areas (MSAs).
All datasets are produced by the U.S. Census Bureau.
We use the \emph{County Business Patterns} (CBP) release \citep{cbp_census}, which provides for each MSA: (i) total employment $E$, (ii) total number of establishments $N$, and (iii) counts of establishments in discrete employment-size bins. Population estimates for MSAs are taken from Census population estimates \citep{census_pop}.

\subsection{Total employment and number of establishments}
\label{sec:NvsE}

We first examine how the total number of establishments $N$ varies with total employment $E$ across MSAs.
These two aggregates are linked by the mean firm size $\langle m \rangle = E/N$.
If the average firm size were approximately constant across MSAs, one would expect a proportionality $N\propto E$.
Systematic deviations from linearity may instead reflect changes in the underlying firm-size distribution with metropolitan scale.

\begin{figure}
    \centering
    \includegraphics[width=0.5\linewidth]{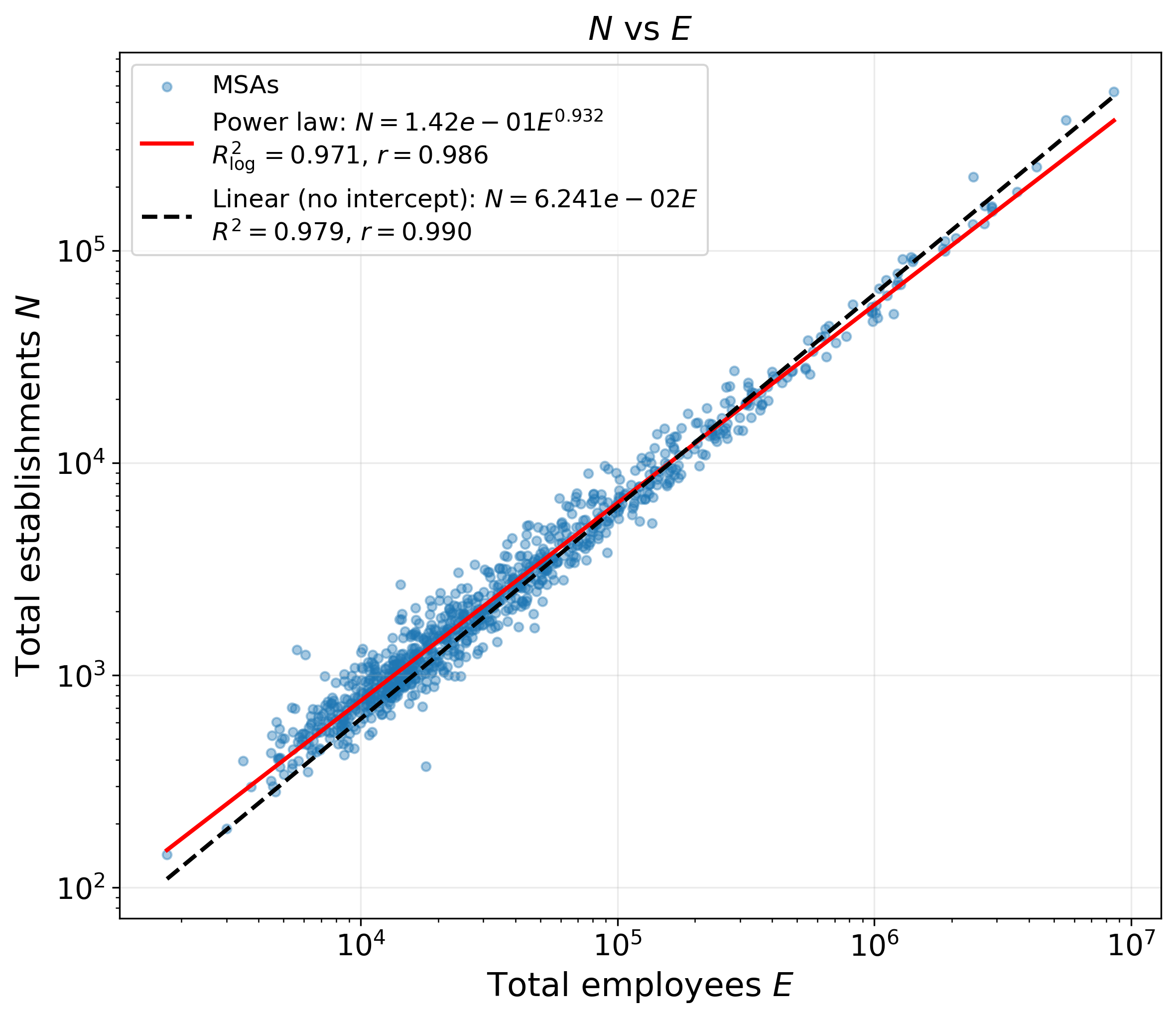}
    \caption{
    Total number of establishments $N$ as a function of total employment $E$ across U.S.\ MSAs.
    Each point is one MSA (only MSAs with $E>0$ and $N>0$ are shown).
    The solid line is a power-law fit $N = a\,E^{\beta}$ obtained from a regression in log--log space.
    The dashed line is a linear fit constrained to pass through the origin, $N = c\,E$, which corresponds to constant mean firm size.
    Both axes are logarithmic; the figure reports correlation and fit-quality diagnostics.
    }
    \label{fig:NvsE}
\end{figure}

Figure~\ref{fig:NvsE} shows $N$ as a function of $E$ for all MSAs with strictly positive values, using logarithmic axes.
We compare two different fits:
(i) a power-law scaling $N = a\,E^{\beta}$ obtained from linear regression in log--log space, and
(ii) a linear relation constrained to pass through the origin, $N = c\,E$, corresponding to constant mean firm size.
To quantify the strength of the relationship, the figure reports the Pearson correlation coefficient $r$ (computed in log space) and the corresponding goodness-of-fit indicators $R^2$.
The power-law fit is very close to linear, with an exponent $\simeq 0.93$, and is nearly indistinguishable from a purely linear regression over the observed range. We therefore conclude that the data are consistent with a linear relationship of the form
\begin{align}
N \simeq \alpha E ,
\end{align}
where $1/\alpha$ corresponds to the average number of employees per establishment. In the following, we adopt this linear approximation in both the analytical model and the numerical experiments, using the empirical value $\alpha \simeq 0.06$.

\subsection{Employment and population}
\label{sec:EvsP}

\begin{figure}
    \centering
    \includegraphics[width=0.6\linewidth]{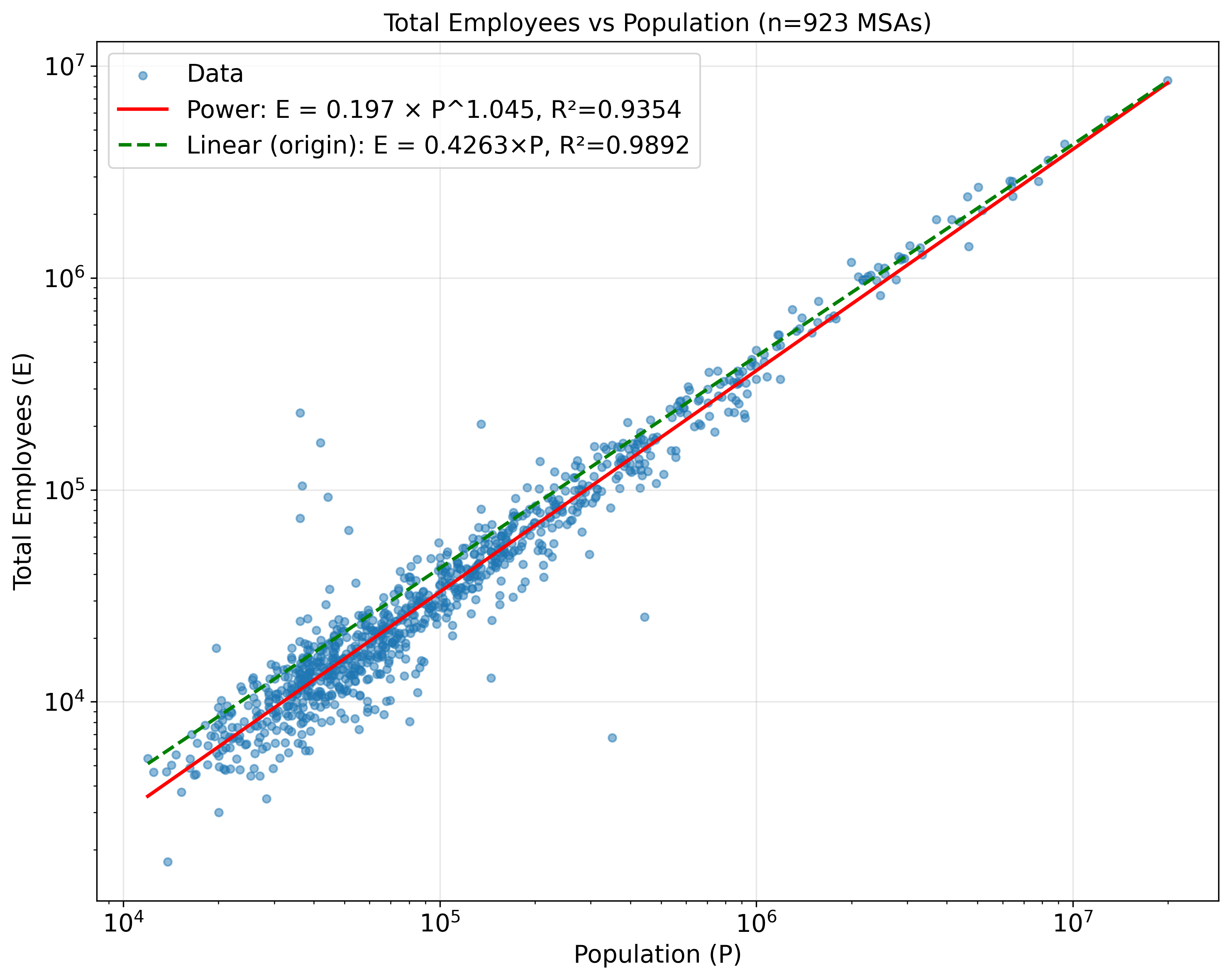}
    \caption{
    Total employment $E$ as a function of MSA population $P$ for MSAs with matched population estimates.
    Each point is one MSA.
    Axes are shown on logarithmic scales.
    (Caption to be updated once the final fitting/diagnostic overlays used in the notebook are fixed and exported.)
    }
    \label{fig:EvsP}
\end{figure}

Although we won't use it in this paper, population provides an additional control for metropolitan size. Once MSA populations are matched to the CBP sample, we can analyze how total employment $E$ scales with population $P$ and compare with extensive expectations (e.g.$E\propto P$).
Figure~\ref{fig:EvsP} reports the resulting relationship.
Here again, the data are consistent with a linear relationship of the form
\begin{align}
E = \eta P ,
\end{align}
with $\eta \simeq 0.43$ for the case of MSAs. The parameter $\eta$ thus corresponds to the fraction of active workers in the total population.

\subsection{Estimation of the Zipf/Pareto tail exponent from binned CBP data}
\label{sec:nu_estimation}

\subsubsection{Method}

For each Metropolitan Statistical Area (MSA), the County Business Patterns (CBP)
dataset reports the number of establishments whose employment size $m$
(number of employees) falls within predefined size classes
\citep{cbp_census}.
Let
\[
[a_i,b_i], \qquad i=1,\dots,B,
\]
denote the boundaries of the employment-size bins, and let $n_i$ be the
corresponding number of establishments in bin $i$, where $B$ is the total
number of bins. All bins are finite except for the top open-ended bin,
denoted $[a_B,\infty)$.

In the CBP tables analyzed here, the open-ended bin corresponds to
establishments with $m \ge 5000$ employees. The complete set of bins is
\begin{align}
& m<5,\nonumber\\
& 5\le m\le 9,\nonumber\\
& 10\le m\le 19,\nonumber\\
& 20\le m\le 49,\nonumber\\
& 50\le m\le 99,\nonumber\\
& 100\le m\le 249,\nonumber\\
& 250\le m\le 499,\nonumber\\
& 500\le m\le 999,\nonumber\\
& 1000\le m\le 1499,\nonumber\\
& 1500\le m\le 2499,\nonumber\\
& 2500\le m\le 4999,\nonumber\\
& m\ge 5000 .
\end{align}

Empirical firm-size distributions very likely deviate from a power law at small sizes. Accordingly, we do not attempt to model the full distribution $P(m)$ over all $m$. Instead, we assume that for sufficiently large firm sizes $m\ge m_{\min}$,
the distribution follows a Pareto (power-law) form,
\begin{equation}
p(m \mid m\ge m_{\min}) = C\,m^{-\nu}, \qquad m \ge m_{\min},
\label{eq:pareto_pdf}
\end{equation}
where $\nu>1$ is the tail exponent and $C$ ensures normalization over the
\emph{conditional} support $[m_{\min},\infty)$.

Crucially, this formulation does \emph{not} impose any assumption on the distribution
of firm sizes below $m_{\min}$. The probability mass below the cutoff,
$q \equiv \mathbb{P}(m\ge m_{\min})$, is left unspecified and is estimated empirically
from the data. The power-law model thus applies only to the conditional tail
distribution $P(m\mid m\ge m_{\min})$.

The associated Zipf (rank--size) exponent is given by (see e.g. \cite{adamic})
\begin{equation}
\gamma \equiv \frac{1}{\nu-1},
\label{eq:gamma_def}
\end{equation}
which yields the standard rank--size scaling $m(r)\propto r^{-\gamma}$ used in the main text.

For $\nu\neq 1$, normalization of \eqref{eq:pareto_pdf} implies
\[
C = (\nu-1)\,m_{\min}^{\nu-1},
\qquad
\mathbb{P}(m\ge x \mid m\ge m_{\min})=
\left(\frac{m_{\min}}{x}\right)^{\nu-1}.
\]

Hence, for a finite bin $[a_i,b_i]$ with $a_i\ge m_{\min}$ and $b_i<\infty$,
the conditional probability mass is
\begin{equation}
P_i(\nu;m_{\min})
= \mathbb{P}(a_i\le m\le b_i \mid m\ge m_{\min})
= \frac{a_i^{1-\nu}-b_i^{1-\nu}}{m_{\min}^{1-\nu}}.
\label{eq:bin_mass_finite}
\end{equation}
For the open-ended bin $[a_B,\infty)$ with $a_B\ge m_{\min}$,
\begin{equation}
P_\infty(\nu;m_{\min})
= \mathbb{P}(m\ge a_B \mid m\ge m_{\min})
= \left(\frac{m_{\min}}{a_B}\right)^{\nu-1}.
\label{eq:bin_mass_open}
\end{equation}
By construction, the probabilities $\{P_b\}$ sum to unity over the retained
tail bins.


For a fixed cutoff $m_{\min}$, we retain only bins with lower boundary
$a_b\ge m_{\min}$ and nonzero counts $n_b>0$.
Let $\mathcal{B}(m_{\min})$ denote this set of tail bins.
Conditioned on $m\ge m_{\min}$, the binned counts follow a multinomial
distribution with probabilities $\{P_b(\nu;m_{\min})\}$.
The corresponding log-likelihood reads
\begin{equation}
\log \mathcal{L}(\nu;m_{\min})
= \sum_{b\in \mathcal{B}(m_{\min})} n_b \log P_b(\nu;m_{\min}),
\label{eq:loglik_binned}
\end{equation}
where $P_b$ is given by Eq.~\ref{eq:bin_mass_finite} (or Eq.~\ref{eq:bin_mass_open} for the last bin).
The maximum-likelihood estimate of the tail exponent is
\begin{equation}
\widehat{\nu}(m_{\min})
= \arg\max_{\nu>1}\,\log \mathcal{L}(\nu;m_{\min}).
\label{eq:mle_nu}
\end{equation}

In practice, we perform a bounded one-dimensional numerical optimization
over $\nu\in[1.1,6]$ and only attempt a fit when at least two bins remain
after thresholding, i.e.\ $|\mathcal{B}(m_{\min})|\ge 2$.
The inferred Zipf exponent then follows from \eqref{eq:gamma_def}:
\begin{equation}
\widehat{\gamma}(m_{\min})
= \frac{1}{\widehat{\nu}(m_{\min})-1}.
\label{eq:gamma_hat}
\end{equation}

\subsubsection{Empirical results}

Using the estimator described above, we compute $\widehat{\nu}$ for each MSA
and convert it to the corresponding Zipf exponent
$\widehat{\gamma}=1/(\widehat{\nu}-1)$.
Since the Pareto assumption is intended to hold only for sufficiently large
establishments, we focus on a tail threshold $m_{\min}=50$ employees
(unless otherwise stated) and report the resulting distribution of exponents
across MSAs.

\begin{figure}
    \centering
    \includegraphics[width=0.6\linewidth]{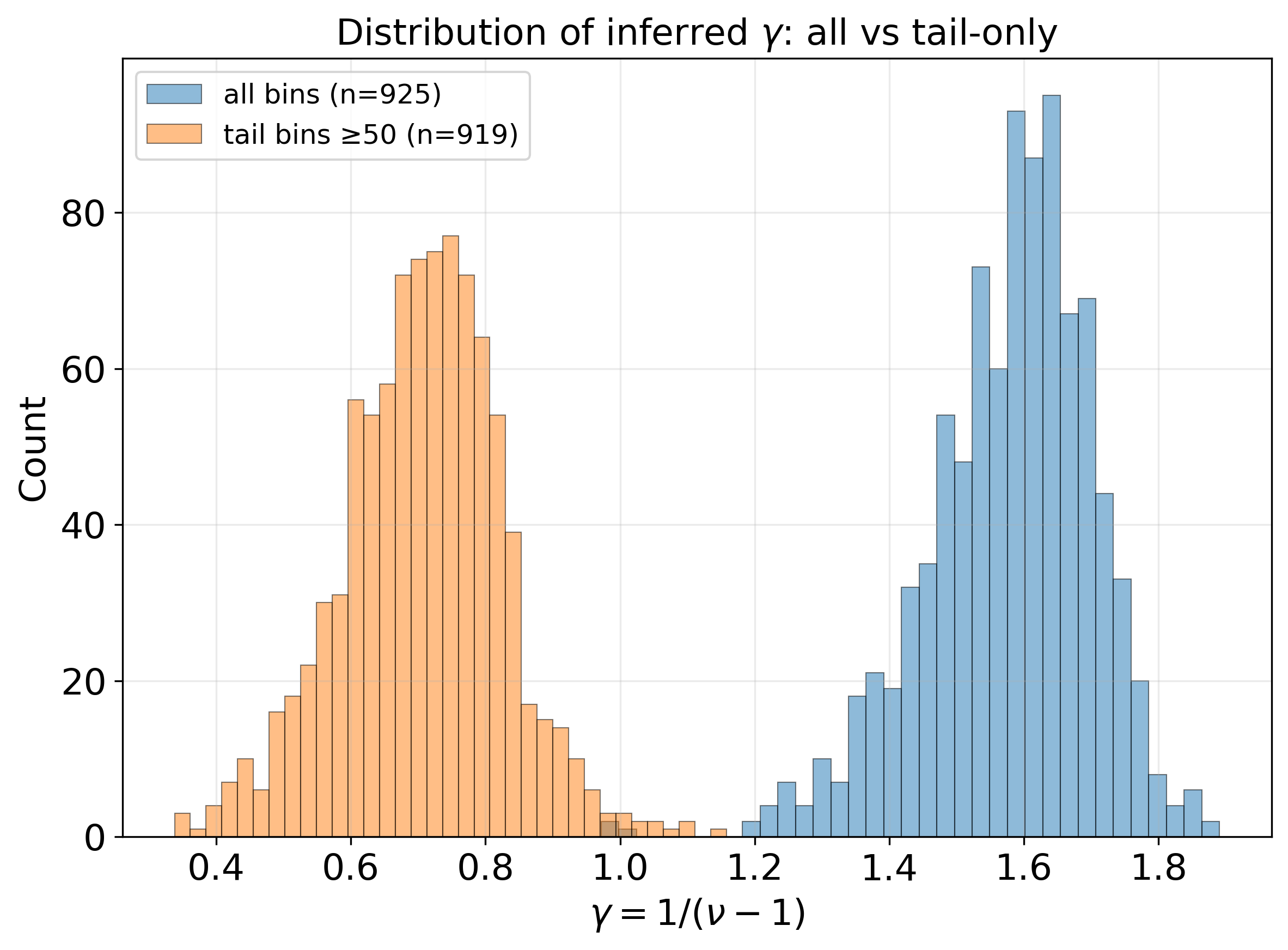}
    \caption{
    Distribution of inferred Zipf exponents $\gamma=1/(\nu-1)$ across U.S.\ MSAs.
    The figure compares estimates obtained using all non-empty bins versus tail-only estimation above a lower cutoff $m_{\min}$ (here $m_{\min}=50$ employees, i.e.\ retaining CBP bins with lower boundary $a_b\ge 50$).
    }
    \label{fig:gamma_all_vs_tail}
\end{figure}

Figure~\ref{fig:gamma_all_vs_tail} compares the inferred Zipf exponents obtained
(i) by fitting all non-empty bins and
(ii) by restricting the estimation to tail bins with lower boundary
$a_b\ge m_{\min}$.
The two procedures yield markedly different results.
When all bins are included, the distribution is narrowly peaked around
$\gamma_{\mathrm{all}}\simeq 1.6$ (corresponding to $\nu_{\mathrm{all}}\simeq 1.64$),
with relatively small dispersion across MSAs.
In contrast, tail-only estimation produces a substantially smaller exponent,
with mean $\gamma_{\mathrm{tail}}\simeq 0.71$
(corresponding to $\nu_{\mathrm{tail}}\simeq 2.46$),
and a noticeably broader distribution. 

To assess the robustness of the inferred tail exponent, we repeated the
estimation procedure for different choices of the lower cutoff $m_{\min}$.
In addition to the baseline value $m_{\min}=50$, we considered thresholds
$m_{\min}=30$ and $m_{\min}=100$ employees, corresponding to progressively
looser and stricter definitions of the tail.

As expected, the inferred exponent exhibits a systematic dependence on
$m_{\min}$.
For $m_{\min}=30$, the average tail exponent across MSAs is
$\langle\gamma_{\mathrm{tail}}\rangle\simeq 1.25$
($\langle\nu_{\mathrm{tail}}\rangle\simeq 1.81$),
reflecting residual influence from the upper part of the distribution body.
Increasing the cutoff to $m_{\min}=50$ yields
$\langle\gamma_{\mathrm{tail}}\rangle\simeq 0.71$
($\langle\nu_{\mathrm{tail}}\rangle\simeq 2.46$),
while a more conservative choice $m_{\min}=100$ further lowers the exponent to
$\langle\gamma_{\mathrm{tail}}\rangle\simeq 0.65$
($\langle\nu_{\mathrm{tail}}\rangle\simeq 2.60$),
at the cost of reducing the number of MSAs with sufficient tail data.

This monotonic decrease of $\gamma_{\mathrm{tail}}$ with increasing
$m_{\min}$ is consistent with a gradual crossover from the body of the
firm-size distribution to its asymptotic tail.
Importantly, once the cutoff is chosen within a reasonable range
($m_{\min}\gtrsim 50$), the inferred exponents stabilize in a narrow interval
$\gamma_{\mathrm{tail}}\approx 0.6$--$0.7$, indicating a robust heavy-tailed
regime for large establishments.
The qualitative conclusions of the paper are insensitive to the precise
choice of $m_{\min}$ within this range.

This systematic shift reflects the fact that the lower part of the firm-size
distribution is probably not well described by a pure power law.
Including small-size bins effectively forces the fit to compromise between
the body and the tail, resulting in artificially shallow Pareto exponents.
Restricting the estimation to the tail isolates the scaling behavior of
large establishments, which is the regime of interest for aggregate
employment concentration and commuting constraints. Importantly, the tail-only exponents remain remarkably stable across MSAs despite large variations in total employment ($E\simeq 1.4\times 10^{5}$ on average, with standard deviation exceeding
the mean by a factor $\sim 3.6$).
This suggests that the heavy-tailed nature of firm sizes is a robust,
city-independent feature of urban economies, while the body of the
distribution primarily reflects local institutional and sectoral effects.

Throughout the main text, we therefore rely on tail-based estimates
$\widehat{\gamma}_{\mathrm{tail}}$, which provide a more faithful
characterization of the asymptotic firm-size heterogeneity relevant
for proximity and accessibility constraints.

\subsubsection{Prediction of the largest establishment size}
\label{sec:m1_prediction}

Given the total employment $E$, the total number of establishments
$N_{\mathrm{est}}$, and an estimated tail exponent $\widehat{\gamma}$,
the model yields a theoretical prediction for the size of the largest
establishment, denoted $m_1^{\mathrm{theo}}$.
This prediction follows from the normalization of the associated
rank--size relation. Assuming a Zipf law for establishment sizes,
\begin{equation}
m(r)=m_1\,r^{-\gamma},
\qquad r=1,\dots,N_{\mathrm{est}},
\end{equation}
the total employment satisfies
\begin{equation}
E=\sum_{r=1}^{N_{\mathrm{est}}} m(r)
= m_1 \sum_{r=1}^{N_{\mathrm{est}}} r^{-\gamma}
\equiv m_1\,H_{N_{\mathrm{est}}}(\gamma),
\end{equation}
where $H_N(\gamma)=\sum_{r=1}^N r^{-\gamma}$ is the generalized harmonic number.
This immediately yields the theoretical prediction
\begin{equation}
m_1^{\mathrm{theo}}=\frac{E}{H_{N_{\mathrm{est}}}(\widehat{\gamma})}.
\label{eq:m1_theo}
\end{equation}

\begin{figure}
    \centering
    \includegraphics[width=0.6\linewidth]{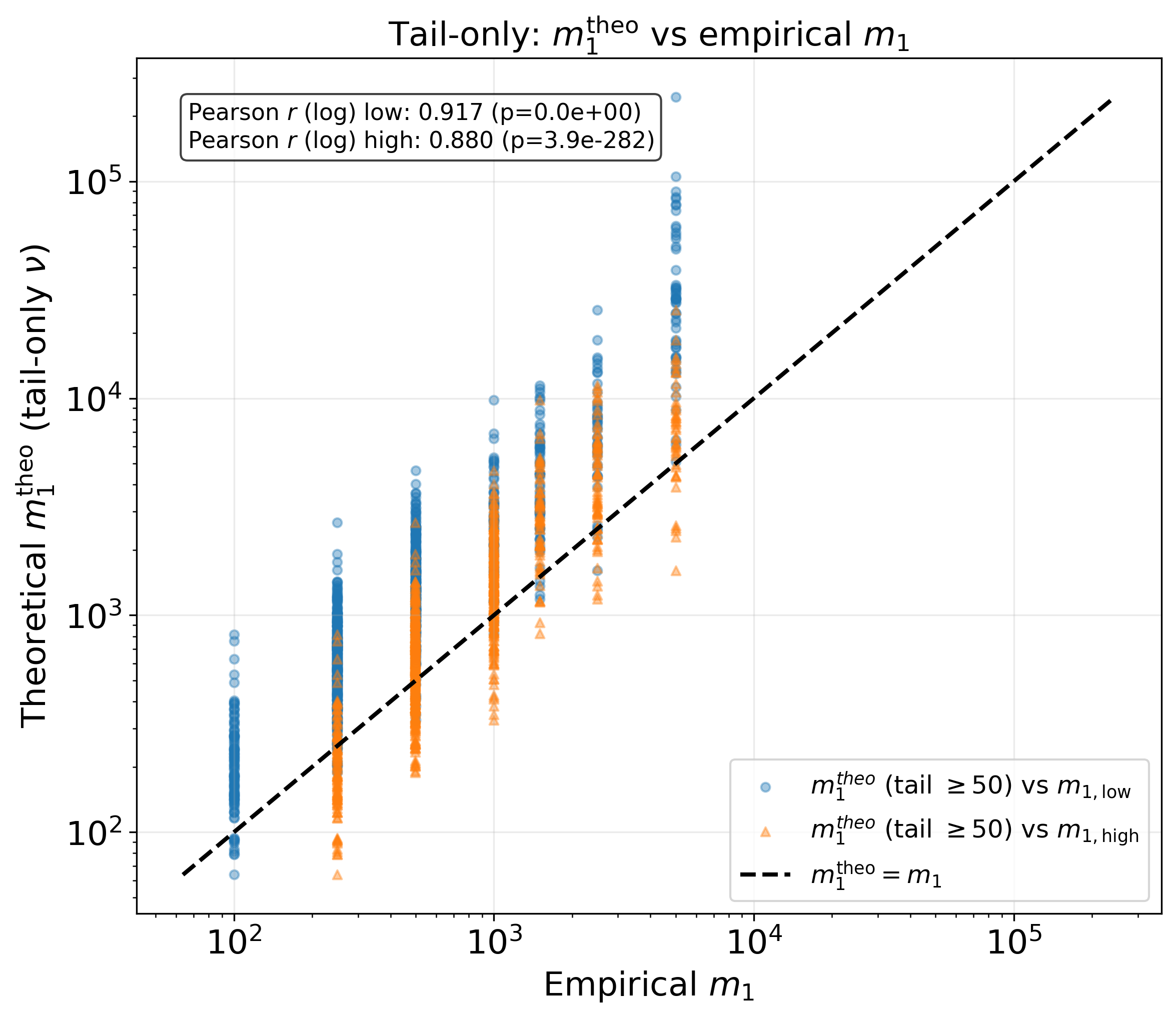}
    \caption{
    Tail-only theory--data comparison for the largest establishment size.
    For each MSA, $m_1^{\mathrm{theo}}$ is computed using the tail-based exponent estimate (retaining bins with $a_b\ge m_{\min}$, here $m_{\min}=50$).
    The empirical largest-establishment size is bracketed by the lower and upper boundaries of the largest non-empty CBP bin, $(m_{1,\mathrm{low}}, m_{1,\mathrm{high}})$.
    The dashed line indicates $m_1^{\mathrm{theo}}=m_1$.
    }
    \label{fig:m1_tail_only}
\end{figure}

We compare this theoretical prediction to an empirical proxy extracted
from the CBP binning. For each MSA, the largest non-empty employment bin provides lower and upper bounds $(m_{1,\mathrm{low}},m_{1,\mathrm{high}})$ on the size of the largest observed establishment.
Because the precise firm size within the top bin is unknown, this comparison
is necessarily approximate and should be interpreted within these bounds.

We focus on the tail-based exponent estimate (with cutoff $m_{\min}=50$),
which is consistent with the modeling assumption that the Pareto form applies
only at large employment sizes.
Figure~\ref{fig:m1_tail_only} shows the comparison between the theoretical
prediction $m_1^{\mathrm{theo}}$ and the empirical bounds
$(m_{1,\mathrm{low}},m_{1,\mathrm{high}})$ across MSAs.
The dashed diagonal indicates perfect agreement,
$m_1^{\mathrm{theo}}=m_1$.

Despite the coarse binning of the empirical data, the theoretical predictions
fall systematically within, or close to, the empirical bounds.
To quantify this agreement, we compute Pearson correlation coefficients
between $\log m_1^{\mathrm{theo}}$ and $\log m_{1,\mathrm{low}}$
(and similarly for $m_{1,\mathrm{high}}$ where available).
For the lower-bound comparison, we find a strong correlation
($r\simeq 0.92$), indicating that the model captures the cross-MSA variability
in the size of the largest establishment remarkably well.
This agreement provides an independent consistency check on both the
tail-based estimation of $\gamma$ and the internal coherence of the model.

\newpage
\clearpage
\section{Theoretical analysis: model definition and feasibility bounds}

\subsection{Model definition}

We consider an idealized city of characteristic radius $R$ and total population $P$, total number of employees $E$, with residents uniformly distributed in space unless stated otherwise. Employment is
provided by $N$ firms, whose sizes are heterogeneous and ranked by decreasing
employment. Firm sizes follow a Zipf distribution, such that the firm of rank $r$ employs $m_r \propto r^{-\gamma}$ workers, where the exponent $\gamma$ controls the degree of economic inequality: $\gamma=0$ corresponds to homogeneous firms, while larger values of $\gamma$ indicate increasingly concentrated employment in a small number of large firms.

More precisely, the idealized city occupies a disk $\Omega=\{\mathbf{r}\in\mathbb{R}^2:\|\mathbf{r}\|\le R\}$
with total employment (and population, assuming full employment) $E$.
Residents are distributed with a radially symmetric density $\rho(r)$, normalized as
\begin{equation}
\int_{\Omega}\rho(\mathbf{r})\,d^2\mathbf{r} \;=\; E.
\end{equation}
Unless stated otherwise, commuting distances are Euclidean.

Employment is provided by $N$ establishments (``firms''), indexed by decreasing size (rank)
$r=1,\dots,N$. Establishment sizes follow a Zipf law
\begin{equation}
m_r \;=\; \frac{m_1}{r^\gamma}, \qquad r=1,\dots,N,
\label{eq:zipf}
\end{equation}
where $\gamma\ge 0$ controls heterogeneity.
The normalization $\sum_{r=1}^N m_r = E$ implies
\begin{equation}
m_1 \;=\; \frac{E}{H_N^{(\gamma)}},
\qquad
H_N^{(\gamma)} \;=\;\sum_{r=1}^N \frac{1}{r^\gamma}.
\label{eq:m1_harmonic}
\end{equation}

The behavior of $H_N(\gamma)$ for large $N$ depends on the value of $\gamma$:
\begin{align}
H_N(\gamma)\sim \begin{cases}
\frac{N^{1-\gamma}}{1-\gamma}\quad &\mathrm{for}\; \gamma<1\\
\log N+\gamma_e \quad & \mathrm{for}\; \gamma=1\\
\zeta(\gamma)+{\cal O}(N^{1-\gamma})\quad & \mathrm{for}\; \gamma>1
\end{cases}
\end{align}

Empirically, the number of establishments scales approximately linearly with total employment:
\begin{equation}
N \;=\; \alpha E,
\label{eq:N_alphaE}
\end{equation}
with $\alpha \approx 0.06$ in our datasets (mean establishment size $\simeq 1/\alpha$).

\subsection{Feasibility problem and definition of $\ell_{\max}$}

The x-minute city constraint requires that every worker can reach their assigned firm within travel distance $\ell$. The theoretical minimum $\ell_{\text{theo}}$ is determined by the largest firm, which must capture a fraction $1/H_N^{(\gamma)}$ of the total population within its catchment area.

Let $\mathbf{x}_r\in\Omega$ be the location of establishment $r$.
A feasible assignment at commuting threshold $L_0$ is a partition of the population into disjoint
catchment regions $A_r\subset\Omega$ such that
\begin{equation}
\int_{A_r}\rho(\mathbf{r})\,d^2\mathbf{r}=m_r,
\qquad
\sup_{\mathbf{r}\in A_r}\|\mathbf{r}-\mathbf{x}_r\|\le L_0
\qquad
\forall r,
\label{eq:feasibility_def}
\end{equation}
and $\cup_{r=1}^N A_r=\Omega$.
For a given configuration, the maximum commute distance is
\begin{equation}
\ell_{\max} \;=\; \max_{r}\;\sup_{\mathbf{r}\in A_r}\|\mathbf{r}-\mathbf{x}_r\|.
\end{equation}
The central question is whether there exist locations $\{\mathbf{x}_r\}$ and an assignment
$\{A_r\}$ such that $\ell_{\max}\le L_0$.

\subsection{A general lower bound from the largest establishment}

A necessary condition for feasibility follows from the largest establishment.
Even with optimal placement of establishment $r=1$ at the density maximum,
its catchment must contain $m_1=E/H_N^{(\gamma)}$ workers within radius $L_0$.
Define the \emph{cumulative population fraction} within distance $\ell$ of the optimal point
(by symmetry, the center for the radial models considered here):
\begin{equation}
F(\ell) \;=\; \frac{1}{E}\int_{\|\mathbf{r}\|\le \ell}\rho(\mathbf{r})\,d^2\mathbf{r}.
\label{eq:W_def}
\end{equation}
Then feasibility requires
\begin{equation}
F(L_0)\;\ge\;\frac{m_1}{E} \;=\; \frac{1}{H_N^{(\gamma)}}.
\label{eq:W_bound}
\end{equation}
Equivalently, a \emph{theoretical feasibility bound} is obtained by solving
\begin{equation}
F(L_0^*) \;=\; \frac{1}{H_N(\gamma)}
\qquad\Rightarrow\qquad
L_0 \;\ge\; L_0^*.
\label{eq:ltheo_general}
\end{equation}
This bound is \emph{necessary} (but not sufficient): even if the largest establishment can be
served within $L_0^*$, other capacity constraints may still force a larger $L_0$.

A complementary (more local) necessary bound uses only the maximum density
$\rho_{\max}=\max_{\mathbf{r}\in\Omega}\rho(\mathbf{r})$:
\begin{align}
\nonumber
m_1 &\;\le\; \int_{\|\mathbf{r}\|\le L_0}\rho(\mathbf{r})\,d^2\mathbf{r}
\;\le\; \rho_{\max}\,\pi L_0^2\\
&\Rightarrow\quad
L_0 \;\ge\; \sqrt{\frac{m_1}{\pi\rho_{\max}}}.
\label{eq:rho_max_bound}
\end{align}
For radially decreasing densities, the bound~\eqref{eq:ltheo_general} is sharper.

\subsection{Uniform density}

We first consider the case of a uniform employment density over a disk of
radius $R$, i.e.
\[
\rho(r)=\frac{E}{\pi R^2}.
\]
In this case, Eq.~\eqref{eq:W_def} yields
\begin{equation}
F(\ell)=\left(\frac{\ell}{R}\right)^2,
\qquad 0\le \ell\le R,
\end{equation}
which simply corresponds to the fraction of area contained within a disk
of radius $\ell$.

Inserting this expression into Eq.~\eqref{eq:ltheo_general} and solving for
the characteristic distance yields
\begin{equation}
L_0^{*\,\mathrm{(unif)}}=\frac{R}{\sqrt{H_N(\gamma)}},
\label{eq:ltheo_uniform}
\end{equation}
where $H_N(\gamma)=\sum_{r=1}^N r^{-\gamma}$ denotes the generalized harmonic
number associated with the Zipf exponent $\gamma$.

Equation~\eqref{eq:ltheo_uniform} highlights how firm-size heterogeneity
controls the minimal characteristic distance through $H_N(\gamma)$.
Several limiting cases are of particular interest:
\begin{itemize}
\item \textbf{Homogeneous establishments} ($\gamma=0$).  
In this case $H_N(0)=N$, leading to
\[
L_0^{*\,\mathrm{(unif)}}=\frac{R}{\sqrt{N}},
\]
which corresponds to the typical spacing of a space-filling tessellation
(e.g.\ optimal hexagonal packing).

\item \textbf{Zipf regime} ($\gamma=1$).  
The harmonic sum diverges logarithmically,
$H_N(1)\simeq \ln N$, yielding
\[
L_0^{*\,\mathrm{(unif)}}\sim \frac{R}{\sqrt{\ln N}}.
\]

\item \textbf{Strong heterogeneity} ($\gamma>1$).  
The harmonic sum converges to the Riemann zeta function,
$H_N(\gamma)\to \zeta(\gamma)$ as $N\to\infty$, implying
\[
L_0^{*\,\mathrm{(unif)}}\to \frac{R}{\sqrt{\zeta(\gamma)}}.
\]
In this regime, the distance threshold becomes independent of the
number of establishments and is controlled solely by the degree of
heterogeneity.
\end{itemize}

\subsection{Exponential radial density}

For centralized cities, we consider an exponential density (a standard assumption for monocentric cities, see for example the original work \citep{clark1951urban})
\begin{equation}
\rho(r) \;=\; \rho_0\,e^{-r/R_d},
\qquad r\ge 0,
\label{eq:rho_exp}
\end{equation}
with decay length $R_d$ and an outer cutoff $R\gg R_d$.
In the limit $R\to\infty$, the normalization gives $E=2\pi\rho_0 R_d^2$ and
\begin{equation}
F(\ell) \;=\; 1-\left(1+\frac{\ell}{R_d}\right)\exp\!\left(-\frac{\ell}{R_d}\right).
\label{eq:W_exp}
\end{equation}
The bound~\eqref{eq:ltheo_general} becomes
\begin{equation}
1-\left(1+x\right)e^{-x} \;=\; \frac{1}{H_N(\gamma)},
\qquad x=\frac{L_0^{*\,\mathrm{(exp)}}}{R_d}.
\label{eq:exp_equation}
\end{equation}
Equivalently,
\begin{equation}
(1+x)e^{-x} \;=\; 1-\frac{1}{H_N(\gamma)}.
\end{equation}
This equation can be solved numerically for $x$.
An explicit closed form uses the Lambert $W$ function:
\begin{equation}
x \;=\; -1 - W\!\left(-\frac{1-\frac{1}{H_N(\gamma)}}{e}\right),
\qquad
L_0^{*\,\mathrm{(exp)}} \;=\; R_d\,x.
\label{eq:lambert_solution}
\end{equation}

For small radii ($x\ll 1$), Eq.~\eqref{eq:W_exp} gives
$F(\ell)\simeq x^2/2$, so the exponential case behaves as
\begin{equation}
L_0^{*\,\mathrm{(exp)}} \;\simeq\; R_d \sqrt{\frac{2}{H_N(\gamma)}}
\qquad \text{(when } 1/H_N(\gamma)\ll 1\text{)}.
\end{equation}
Compared at fixed $H_N(\gamma)$, the required $L_0^{*}$ depends on the
chosen macroscopic scale ($R$ vs $R_d$); thus, uniform vs exponential should be compared
using consistent city-size conventions (e.g., fixing the same outer radius or the same mean density).

\subsection{Interpretation}

Equations~\eqref{eq:ltheo_general}--\eqref{eq:lambert_solution} show that firm-size inequality
enters the feasibility condition only through the fraction $m_1/E=1/H_N(\gamma)$ that must
be served by the largest establishment. For $\gamma<1$ and $N=\alpha E$, one has
$H_N(\gamma)\sim N^{1-\gamma}/(1-\gamma)$, hence $m_1\sim E/N^{1-\gamma}\propto E^\gamma$,
so the constraint induced by the head of the distribution strengthens with system size.

Summary Table:

\begin{table}[h]
\centering
\begin{tabular}{lcc}
\hline
\textbf{Property} & \textbf{Uniform Density} & \textbf{Exponential Density} \\
\hline
Density profile & $\rho(r) = \rho_0$ & $\rho(r) = \rho_0 e^{-r/R}$ \\
Weight function & $F(x) = x^2$ & $F(x) = 1 - (1+x)e^{-x}$ \\
Theoretical bound & $L_0^* = R/\sqrt{H_N}$ & Solve numerically \\
Small $L_0^*$ behavior & $L_0^* \propto 1/\sqrt{H_N}$ & $L_0^* \propto \sqrt{2/H_N}$ \\
\hline
\end{tabular}
\caption{Comparison of theoretical bounds for uniform and exponential population densities.}
\end{table}

\newpage
\clearpage
\section{Dependence on capacity tolerance}
\label{sec:effect_delta}

Our model allows firms to operate with flexible capacity: firm~$i$ with target size~$m_i$ can employ between $(1-\delta)m_i$ and $(1+\delta)m_i$ workers, where $\delta \in [0,1]$ is the tolerance parameter. This flexibility captures realistic features of labor markets, including part-time employment, seasonal fluctuations, and firms operating below maximum capacity. Here we examine how~$\delta$ affects the optimized accessibility ratio across different firm-size distributions parametrized by the Zipf exponent~$\gamma$.


The theoretical lower bound $\ell_{\mathrm{theo}}^{(0)} = R/\sqrt{H_N^{(\gamma)}}$ assumes \emph{strict} capacity constraints ($\delta=0$), where each firm must employ exactly~$m_i$ workers. With flexible capacity, the constraint is relaxed: the largest firm need only recruit its \emph{minimum} required workforce $(1-\delta)\,m_1 = (1-\delta)\,P/H_N^{(\gamma)}$. When optimally placed at the population density maximum, this requirement becomes:
\begin{equation}
  F\!\left(\frac{\ell}{R}\right) \;\geq\; \frac{1-\delta}{H_N^{(\gamma)}}\,,
\end{equation}
where $F(x)$ is the cumulative population fraction within dimensionless radius~$x$. The flexibility-corrected theoretical bound is therefore:
\begin{equation}
  \ell_{\mathrm{theo}}^{(\delta)} \;=\; R \cdot F^{-1}\!\left(\frac{1-\delta}{H_N^{(\gamma)}}\right).
  \label{eq:l_theo_delta}
\end{equation}
For uniform population density, $F(x) = x^2$, yielding the explicit formula:
\begin{equation}
  \ell_{\mathrm{theo}}^{(\delta)}
  = R\,\sqrt{\frac{1-\delta}{H_N^{(\gamma)}}}
  = \sqrt{1-\delta}\;\cdot\;\ell_{\mathrm{theo}}^{(0)}\,.
  \label{eq:l_theo_delta_uniform}
\end{equation}
For $\delta = 30\%$, this gives $\ell_{\mathrm{theo}}^{(0.3)} = \sqrt{0.7}\;\ell_{\mathrm{theo}}^{(0)} \approx 0.84\;\ell_{\mathrm{theo}}^{(0)}$, meaning that a $30\%$ flexibility margin reduces the minimum feasible travel distance by approximately $16\%$ in the uniform-density case ($1 - \sqrt{0.7} \approx 0.16$). For exponential population density, $F(x) = 1 - (1+x)\,e^{-x}$, and Eq.~\eqref{eq:l_theo_delta} must be inverted numerically.

\newpage
\clearpage
\section{Numerical method}

\subsection{General approach}

The $x$-minute city feasibility problem can be formulated as:

\begin{quote}
\textit{Given parameters $(E, R, N, \gamma, m_{\min})$, find firm locations $\{\mathbf{x}_j\}_{j=1}^N$ and an assignment $c: \{1, \ldots, E\} \to \{1, \ldots, N\}$ such that:}
\begin{enumerate}
    \item Each firm $j$ is assigned exactly $m_j$ employees.
    \item The maximum commute distance $\ell_{\max} \leq L_0$.
\end{enumerate}
\end{quote}

Equivalently, we seek to minimize $\ell_{\max}$ over firm locations:
\begin{equation}
    \ell_{\max} = \min_{\{\mathbf{x}_j\}} \left[ \min_{c} \max_i d(i, c(i)) \right],
\end{equation}
subject to capacity constraints. The feasibility threshold $L_0^*$ is the smallest $L_0$ for which a valid configuration exists.

\subsubsection{Goal and outputs}

The numerical objective is to estimate, for given parameters $(E,R,\gamma,\alpha)$ and a chosen
population density model, the minimal achievable maximum commuting distance
\begin{equation}
\ell_{\max}(\gamma) \;=\;
\min_{\{\mathbf{x}_r\}\subset\Omega}\;
\left[
\min_{\{A_r\}}
\max_{r}\sup_{\mathbf{r}\in A_r}\|\mathbf{r}-\mathbf{x}_r\|
\right],
\end{equation}
under the capacity constraints $\int_{A_r}\rho(\mathbf{r})\,d^2\mathbf{r}=m_r$.
We compare $\ell_{\max}(\gamma)$ to the theoretical lower bound $\ell_{\mathrm{theo}}(\gamma)$
from Sec.~\ref{eq:ltheo_general}, and we build feasibility curves (or ``phase diagrams'')
in the $(\gamma,L_0)$ plane by declaring the system feasible whenever $\ell_{\max}^*(\gamma)\le L_0$.

\subsubsection{Discretization of the population}

We represent the population by $E$ points $\{\mathbf{r}_i\}_{i=1}^E$ sampled independently
from the prescribed density $\rho(r)$ within the disk of radius $R$.
In practice, we use moderate sizes (typically $E\sim 10^4$) to keep the optimization tractable,
and we verify robustness by repeating runs with different random seeds.

\subsubsection{Establishment sizes and empirical scaling $N=\alpha E$}

For each run and each Zipf exponent $\gamma$, we set the number of establishments via
\begin{equation}
N = \lfloor \alpha E \rceil,
\end{equation}
and define capacities $\{m_r\}_{r=1}^N$ by the Zipf law~\eqref{eq:zipf} with normalization~\eqref{eq:m1_harmonic}.
Capacities are then converted to integers that sum exactly to $E$ by rounding and distributing the remainder.
Optionally, we allow a tolerance window $m_r(1\pm \delta)$ to account for finite-sampling fluctuations
in the discretized population (we report $\delta$ explicitly whenever used).


\subsection{Sensitivity to the greedy assignment order}

For fixed establishment positions, the worker--establishment assignment is performed using a capacitated greedy heuristic. In the main implementation, all distances between workers and establishments are first computed. Workers are then sorted by their distance to the nearest establishment, and are processed in increasing order of this distance. Each worker is assigned to the nearest establishment with remaining capacity. Thus the rule is a nearest-feasible-establishment assignment, subject to capacities, rather than a simple nearest-establishment assignment.

Because this procedure is sequential, the processing order can affect the final assignment and hence the measured value of $\ell_{\max}$. To quantify this effect, we repeated the full simulated-annealing optimization with the same parameters,
\[
P=1000,\qquad N=60,\qquad \gamma=1.5,\qquad R=1,
\]
but with different assignment orders used inside the objective function. We compared the original nearest-first ordering, a farthest-first ordering, and 20 fixed random worker orderings. In the random-order tests, the random permutation was fixed during each simulated-annealing run, so that the optimization objective remained deterministic within a run.

For the original nearest-first rule, simulated annealing reduced $\ell_{\max}$ from $1.493$ to $0.669$, corresponding to an improvement of $55\%$. The farthest-first rule gave a very similar result, reducing $\ell_{\max}$ from $1.464$ to $0.680$, corresponding to an improvement of $53.5\%$. For the 20 fixed random orderings, the initial values were
\[
\ell_{\max}^{\rm init}=1.484\pm0.008,
\]
and the final values were
\[
\ell_{\max}^{\rm final}=0.923\pm0.043,
\]
with a range from $0.846$ to $0.992$. The corresponding mean improvement was $37.8\%$, with values ranging from $33.2\%$ to $43.3\%$.

These tests show that the numerical value of $\ell_{\max}$ depends on the assignment order used by the greedy procedure. In particular, the original nearest-first ordering gives a lower final value than typical fixed random orderings. However, the qualitative conclusion is robust: for all assignment orders tested, simulated annealing substantially reduces the maximum commuting distance. The numerical configurations should therefore be interpreted as feasible heuristic solutions obtained under a specified capacitated greedy assignment rule, not as proofs of the global optimum of the capacitated assignment problem.

\subsection{Optimization of establishment locations}

To minimize the maximum commuting distance $\ell_{\max}$ over establishment locations, we
use a simulated annealing (SA) heuristic, which is well suited for high-dimensional problems with many local minima.
The objective function is defined as the maximum Euclidean distance between any worker
and the establishment to which they are assigned, given the firm capacities and the
current spatial configuration.

Each SA run starts from an initial configuration in which establishments are placed on a
regular, approximately hexagonal lattice within the disk of radius $R$, ensuring a
homogeneous initial coverage of space. At each Monte Carlo step, a single establishment
is selected at random and displaced by a small random vector drawn from an isotropic
distribution with a fixed step size. Proposed moves that place the establishment outside
the disk are rejected.

For a proposed displacement, the change in the objective function,
$\Delta \ell_{\max}$, is computed by recomputing the worker--establishment assignment and
the resulting maximum commuting distance. Moves that decrease $\ell_{\max}$ are always
accepted, while moves that increase it are accepted with probability
$\exp(-\Delta \ell_{\max}/T)$, where $T$ is the current annealing temperature (Metropolis
criterion). The temperature is gradually reduced according to a prescribed cooling
schedule until the system becomes effectively frozen and no further improvements are
observed.

To reduce sensitivity to initial conditions and stochastic fluctuations, the SA
procedure is repeated for multiple independent initial seeds and random number streams.
Among all runs, we retain the configuration yielding the smallest value of
$\ell_{\max}$, which is reported as the optimized solution for the given parameter set.

\subsection{Capacity tolerance $\delta$ and feasibility relaxation}
\label{sec:delta_tolerance}

In the discrete formulation, each establishment $r$ must be assigned exactly $m_r$
workers, where the target capacities $\{m_r\}$ follow the Zipf law
(Eq.~\eqref{eq:zipf}) and are converted to integers so that $\sum_{r=1}^N m_r = E$.
For a fixed set of establishment locations $\{\mathbf{x}_r\}$, enforcing the strict
capacity constraints can lead to artificially large $\ell_{\max}$ in finite samples:
even when a ``nearly feasible'' spatial configuration exists, a small mismatch between
local worker availability and discrete integer capacities can force some individuals to
be assigned far away to satisfy the exact counts.

To control for this finite-size/discretization effect, we introduce a \emph{capacity
tolerance} parameter $\delta\ge 0$ that relaxes the hard upper bound on the number of
workers that an establishment can accept. Concretely, in the assignment step we replace
the capacity constraint
\begin{equation}
n_r \le m_r
\end{equation}
by
\begin{equation}
n_r \le m_r^{\max}(\delta)
\;\equiv\;
\left\lceil (1+\delta)\,m_r \right\rceil,
\label{eq:delta_capacity}
\end{equation}
where $n_r$ is the number of workers assigned to establishment $r$ and
$\lceil\cdot\rceil$ denotes the ceiling function. The parameter $\delta$ therefore
controls the allowed relative \emph{overfill} of each establishment, with $\delta=0$
recovering strict capacities. Because the integer rounding in
Eq.~\eqref{eq:delta_capacity} can change the total available capacity, we ensure that
the relaxed total capacity is at least $E$ (otherwise no complete assignment can exist);
in practice we increase all $m_r^{\max}$ by a common multiplicative factor of order
$1+\varepsilon$ whenever $\sum_r m_r^{\max}(\delta)<E$.

Operationally, we use $\delta$ only during the assignment subproblem: given
$\{\mathbf{x}_r\}$, each worker is greedily assigned to the nearest establishment with
remaining relaxed capacity, and the resulting maximum distance defines $\ell_{\max}$ for
that configuration. Simulated annealing then minimizes this relaxed objective over
establishment locations.

The tolerance $\delta$ has two interpretations:
(i) it reduces spurious infeasibility that originates from discrete sampling of a
continuous density field and integer capacities; and
(ii) it can be viewed as a coarse-grained representation of real-world flexibility,
where establishments can absorb modest deviations from target headcounts (through
temporary overstaffing, part-time work, multi-site allocation, or measurement noise in
the empirical size distribution).

In the main numerical results we report $\ell_{\max}(\gamma,\delta)$ for a fixed
$\delta$ (typically $\delta\simeq 0.3$) and show that qualitative trends with $\gamma$
are robust. Importantly, because $\delta>0$ relaxes constraints, $\ell_{\max}(\gamma,\delta)$
is decreasing with $\delta$; hence the strict-capacity case $\delta=0$ provides the
most conservative feasibility assessment, while $\delta>0$ yields an lower estimate of
realistic feasibility when moderate capacity fluctuations are allowed.

\subsection{Numerical Approach: Simulated Annealing}

We employ simulated annealing to optimize firm locations~\citep{kirkpatrick1983}. The algorithm proceeds as follows:

\begin{algorithm}[H]
\caption{Simulated Annealing for Firm Location Optimization}
\KwIn{Initial temperature $T_0$, cooling rate $\alpha \approx 0.95$, minimum temperature $T_{\min}$}
\KwOut{Optimized positions $\{\mathbf{x}_j\}$ and $\ell_{\max}$}
Initialize random firm positions $\{\mathbf{x}_j\}$ within the disk\;
$T \gets T_0$\;
\While{$T > T_{\min}$}{
    \For{$k = 1$ \KwTo $n_{\mathrm{steps}}$}{
        Perturb one firm position: $\mathbf{x}_j \gets \mathbf{x}_j + \boldsymbol{\delta}$\;
        Compute assignment $c$ (nearest-firm or capacitated)\;
        Compute new cost $\mathcal{E}' = \ell_{\max}$\;
        $\Delta \mathcal{E} \gets \mathcal{E}' - \mathcal{E}$\;
        \eIf{$\Delta \mathcal{E} < 0$ \textbf{or} $\mathrm{rand}() < e^{-\Delta \mathcal{E}/T}$}{
            Accept move\;
        }{
            Reject move\;
        }
    }
    $T \gets \alpha T$\;
}
\Return optimized positions $\{\mathbf{x}_j\}$ and $\ell_{\max}$\;
\end{algorithm}

The primary cost function is the maximum commute distance:
\begin{equation}
    {\cal E} = \ell_{\max} = \max_i d(i, c(i)).
\end{equation}

To map the feasibility landscape, we perform a systematic scan:
\begin{enumerate}
    \item For each $\gamma \in [0.5, 2.0]$ (discretized)
    \item Run simulated annealing to find $\ell_{\max}(N, \gamma)$
    \item Record $(N, \gamma, \ell_{\max})$
\end{enumerate}

\newpage
\clearpage
\section{Case study: Paris}
\label{sec:SI_paris}

In this section, we detail the estimation of the firm-size exponent $\gamma$ used in the Paris case study. Direct establishment-size data for the administrative city of Paris are not available at a sufficiently fine resolution.
We therefore rely on publicly available data for the Paris urban area \citep{insee_paris_navetteurs,insee_2023}, which provides the closest empirical proxy for the distribution of economic activity at this scale.

The dataset reports the number of establishments grouped into four employment-size classes:
1--9, 10--19, 20--49, and 50 or more employees.
To estimate the exponent $\gamma$, we assume that firm sizes follow a rank--size (Zipf-like) distribution,
\begin{equation}
m(r) \sim r^{-\gamma},
\end{equation}
which implies a complementary cumulative distribution of the form
$P(m \geq x) \sim x^{-1/\gamma}$.
Under this assumption, the expected fraction of establishments in each size class can be computed analytically.
The exponent $\gamma$ is then estimated by fitting these predicted fractions to the empirical bin counts using a multinomial likelihood (see section I.C).

\begin{figure}[b]
  \centering
  \includegraphics[width=0.7\linewidth]{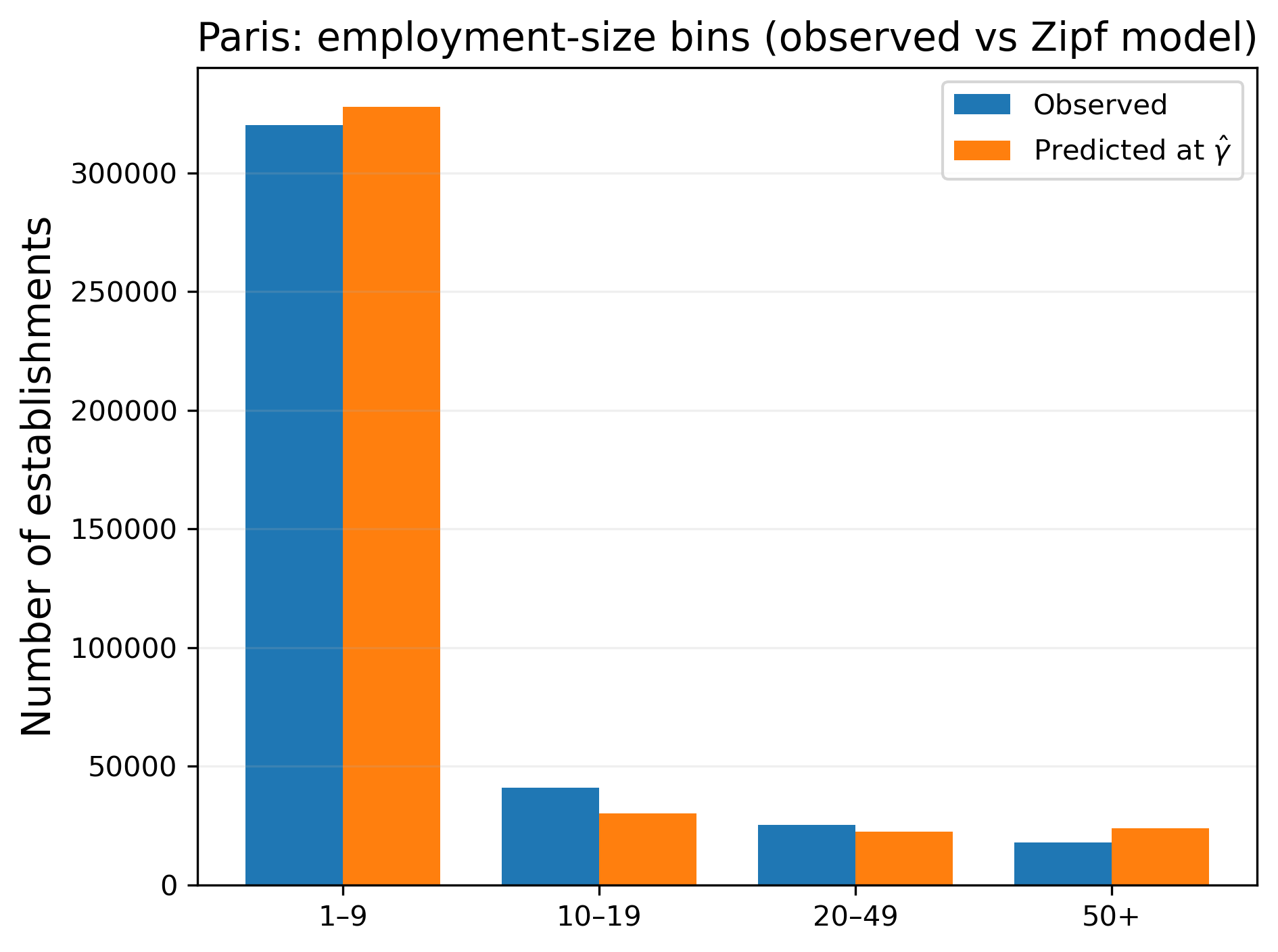}
  \caption{
  Empirical employment-size bins for the Paris urban area compared with a rank--size model fitted with exponent $\hat\gamma\simeq1.38$.
  Bars show the number of establishments in four employment-size classes (1--9, 10--19, 20--49, and 50+ employees), using INSEE data.
  The fitted model captures the pronounced heterogeneity of firm sizes despite the coarse binning.
  }
  \label{fig:gamma_paris}
\end{figure}

Figure~\ref{fig:gamma_paris} compares the empirical employment-size bins with the best-fit theoretical distribution.
The fitted value $\hat\gamma \simeq 1.38$ reflects the strong heterogeneity of firm sizes in the Paris metropolitan area,
with a large number of small establishments coexisting with a long tail of large employers.
Because the data are coarsely binned, this estimate should be interpreted as an effective exponent rather than a precise universal value.
For this reason, $\gamma$ is treated as a control parameter in the main analysis, and a broad range of values is explored to assess the robustness of the feasibility constraints on proximity.

\newpage

\bibliography{ref_xmc}
\bibliographystyle{dcu}